\documentclass[]{aa}  
\usepackage{graphicx}
\usepackage{amsmath}
\usepackage{wasysym}
\usepackage{mathtools} 
\usepackage{booktabs} 
 \usepackage{rotating,multirow} 
\usepackage[USenglish]{isodate}
\cleanlookdateon 

\usepackage{verbatim}

\usepackage{graphicx}

\usepackage{natbib,twoopt}
\usepackage[breaklinks=true]{hyperref} 
\bibpunct{(}{)}{;}{a}{}{,} 
\makeatletter

\usepackage{color}
\usepackage{soul}

\begin{document} 

\title{Rotational evolution of slow-rotator sequence stars}
\titlerunning{Rotational evolution of slow-rotator sequence stars}

\author{
A.~C. Lanzafame\inst{1,2}\fnmsep\thanks{Visiting scientist at the Leibniz-Institut f\"ur Astrophysik Potsdam (AIP)} 
\and F. Spada\inst{3} 
}

\authorrunning{A.~C. Lanzafame \& F. Spada}

\offprints{A.~C. Lanzafame \\ \email{a.lanzafame@unict.it}}

\institute{
Universit\`a di Catania, Dipartimento di Fisica e Astronomia, Sezione Astrofisica, Via S. Sofia 78, I-95123 Catania, Italy \\
 \email{Alessandro.Lanzafame@oact.inaf.it}
\and
INAF-Osservatorio Astrofisico di Catania, Via S. Sofia 78, I-95123 Catania, Italy
\and
Leibniz-Institut f\"ur Astrophysik Potsdam (AIP), An der Sternwarte 16, D-14482, Potsdam, Germany
}

\date{Received 17 June 2015 / Accepted 16 September 2015 }

\abstract{
The observed relationship between mass, age and rotation in open clusters shows the progressive development of a slow-rotator sequence among stars possessing a radiative interior and a convective envelope during their pre-main sequence and main-sequence evolution. After 0.6 Gyr, most cluster members of this type have settled on this sequence.
}{
The observed clustering on this sequence suggests that it corresponds to some equilibrium or asymptotic condition that still lacks a complete theoretical interpretation, and which is crucial to our understanding of the stellar angular momentum evolution.
}{
We couple a rotational evolution model, which takes internal differential rotation into account, with classical and new proposals for the wind braking law, and fit models to the data using a Monte Carlo Markov Chain (MCMC) method tailored to the problem at hand.
We explore to what extent these models are able to reproduce the mass and time dependence of the stellar rotational evolution on the slow-rotator sequence.
}{
The description of the evolution of the slow-rotator sequence requires taking the transfer of angular momentum from the radiative core to the convective envelope into account.
We find that, in the mass range $0.85$--$1.10$ $M_{\odot}$, the core-envelope coupling timescale for stars in the slow-rotator sequence scales as $M^{-7.28}$.
Quasi-solid body rotation is achieved only after $1$--$2$ Gyr, depending on stellar mass, which implies that observing small deviations from the Skumanich law ($P \propto \sqrt{t}$) would require period data of older open clusters than is available to date. 
The observed evolution in the $0.1$--$2.5$ Gyr age range and in the $0.85$--$1.10$ $M_{\odot}$ mass range is best reproduced by assuming an empirical mass dependence of the wind angular momentum loss proportional to the convective turnover timescale and to the stellar moment of inertia. 
Period isochrones based on our MCMC fit provide a tool for inferring stellar ages of solar-like main-sequence stars from their mass and rotation period that is largely independent of the wind braking model adopted.
These effectively represent gyro-chronology relationships that take the physics of the two-zone model for the stellar angular momentum evolution into account.
}
{}

\keywords{Stars: rotation -- Stars: evolution -- Stars: late-type -- open clusters and associations: general -- open clusters and associations: individual: 
\object{Pleiades}, 
\object{M34}, 
\object{M35}, 
\object{M37}, 
\object{Hyades}, 
\object{Praesepe}, 
\object{Coma Berenices}
\object{NGC6811}
\object{NGC6819}
}

\maketitle

\section{Introduction}

In the past decade, stellar rotation has received a great deal of attention, both because of its promising potential as a precise stellar age estimator \citep{Barnes:2007,Barnes:2010,Meibom_etal:2015} and the large amount of good quality data increasingly available, either as a byproduct of planetary transit searches or from dedicated stellar rotation observational programs \citep[see, e.g. ][]{Herbst_Mundt:2005, Hodgkin_etal:2006, vonBraun_etal:2005, Messina:2007,  Messina_etal:2008, Messina_etal:2010}. 
Rotation periods of young stars can in principle be derived with a precision of 10$^{-4}$ using photometric monitoring, with surface differential rotation being probably the most important limiting factor for individual old stars \citep[see, e.g.][]{Epstein_Pinsonneault:2014}.

The realization by \cite{Barnes:2003} of the existence of a well-defined sequence of slowly rotating stars in the young open clusters color-period diagram has provided a criterion for establishing an empirical relationship between mass, age and rotation, which can be used to infer the age of stars that are sufficiently old to belong to that sequence.
\cite{Barnes:2003} effectively disentangled the convergence of stellar periods towards a unique slow-rotator sequence, seen as a progressive reduction of the scatter that is due to the presence of fast rotators (from very young clusters to the Hyades), from the rotational evolution of stars in this sequence, which dominates the color-period diagram from the Hyades onwards.
Subsequently, \cite{Barnes:2007} provided a calibration of his rotation-mass-age relationship using the observed rotation-mass distribution in open clusters and the solar age, rotation period and $(B-V)$.
More recently, \cite{Barnes_Kim:2010} provided evidence of a connection between the empirical rotational period functional dependence on $(B-V)$ and the convective turnover timescale $\tau$. 
Based on this, \cite{Barnes:2010} derived a nonlinear model formulated in terms of the Rossby number ($Ro = P / \tau$) and two dimensionless empirical constants, which describes the rotational evolution towards the slow-rotator sequence.

In this work we focus on the rotational evolution of stars already on the slow-rotator sequence and study how well the current models are able to reproduce such evolution.
We focus on the age interval between $0.1$ and $2.5$ Gyr, for which data with the richest details is available, providing strong constraints on the models.
The observed clustering on this sequence in the mass-age-period space suggests that it corresponds to some equilibrium or asymptotic condition that still lacks a complete theoretical interpretation and may be crucial to our understanding of the stellar angular momentum evolution.

Data that has become available in recent years has revealed an evolutionary behavior of the slow-rotator sequence that seems difficult to reconcile with the original \cite{Barnes:2003} idea of a factorization of the time and mass dependence.
Notably, in their study of rotation in \object{M35} and \object{M34}, \cite{Meibom_etal:2009,Meibom_etal:2011} found that while G-type stars on the slow-rotator sequence spin down with a rate close to the Skumanich $P \propto \sqrt{t}$ law, K-type stars spin down more slowly.  

On the theoretical side, recent work concentrated on improving the description of the magnetic braking due to coupling between the stellar wind and magnetic fields generated by an internal stellar dynamo.
\cite{Matt_etal:2012a} performed two-dimensional axisymmetric magnetohydrodynamic simulations to compute steady-state solutions for solar-like stellar winds from rotating stars with dipolar magnetic fields, from which they derived a semi-analytic formulation for the external torque on the star.
\cite{Gallet_Bouvier:2013} derived the angular momentum loss rate using the \cite{Matt_etal:2012a} equation for the Alfv\'en radius together with a dynamo prescription that relates the stellar magnetic field to stellar rotation, as well as a wind prescription that relates the mass-loss rate to the stellar angular velocity, obtained from current theory and calibrated to the present-day Sun.
Given the current uncertainties on the stellar magnetic fields and on the wind mass-loss, however, \cite{Matt_etal:2015} derived a scaling for the dependence of the stellar wind braking on Rossby number and an empirically-derived scaling with stellar mass and radius.

In this work, we include these various wind braking laws in the classical two-zone description of the stellar rotational evolution and investigate to what extent they reproduce the slow-rotator sequence evolution. 
The models are fitted to the data by means of a Monte Carlo Markov Chain (MCMC) method tailored to the case at hand.
Using the parameters derived by the MCMC model fitting to the data, we derive period isochrones and tracks from 0.1 to 5\,Gyr that are largely independent of the specific wind braking law adopted.
These effectively represent gyro-chronology relationships that take the effects of differential rotation in the period evolution into account.

The paper is organized as follows.
In Sect.\,\ref{sec:data} we discuss the data used in the present analysis.
In Sect.\,\ref{sec:empirical} we present an empirical description of the slow-rotator sequence, which also illustrates the current state-of-the-art, together with a novel non-parametric empirical fit of its evolution.
In Sect.\,\ref{sec:tzm_fitting} we present the MCMC fitting method of two-zone rotational evolution models with different wind braking prescriptions. 
In Sect.\,\ref{sec:results} the results of the MCMC fitting to the data are presented and discussed.
The conclusions are in Sect.\,\ref{sec:conclusions}.

\section{Data}
\label{sec:data}

\begin{table*}
   \centering
      \caption{Clusters data used in this work.}
      \label{tab:data}
      \begin{tabular}{@{}lrrl@{}}
      \hline
Name     &  Age (Myr) & references \\
\hline
Pleiades       & 120 & \cite{Hartman_etal:2010} \\  
M35            & 150 & \cite{Meibom_etal:2009} \\
M34            & 220 & \cite{Meibom_etal:2011} \\ 
M37            & 550 & \cite{Hartman_etal:2009} \\ 
Hyades         & 600 & \cite{Delorme_etal:2011} \\ 
Praesepe       & 600 & \cite{Delorme_etal:2011} \\ 
Coma Berenices & 600 & \cite{CollierCameron_etal:2009} \\
NGC6811        & 1000 & \cite{Meibom_etal:2011,Janes_etal:2013}\\
NGC6819        & 2500 & \cite{Meibom_etal:2015}\\
\hline
      \end{tabular}
\end{table*}

For this work we use literature data of stars in open clusters that show a well-defined slow-rotator sequence, easily identifiable by eye as an over-density in the period-color diagram, also known in the literature as the ``I-sequence'' \citep{Barnes:2003} or ``ridge'' \citep{Kovacs:2015}.
The age at which this sequence begins to form is still rather undetermined, but it is evident that this is already well defined at the age of the Pleiades and present in all older open clusters.
A summary is reported in Table\,\ref{tab:data}.
The data set includes clusters from 0.1 to 2.5 Gyr for which $(B-V)$ measurements in the relevant range are readily available.
The age interval is sampled in an optimal way, in the sense that data for open clusters not considered here would not improve the age sampling in a significant way.
Since our method relies on an empirical fitting of the slow-rotator sequence to start with (see Sect.\,\ref{sec:empirical}), data for field stars are not considered.
Young loose stellar associations are also not considered in this work.

Rotational periods for the \object{Pleiades} are taken from \cite{Hartman_etal:2010}.
This data set includes 241 single stars in common with the compilation of probable Pleiades members by \cite{Stauffer_etal:2007}, which provides near- and mid-infrared photometry together with a compilation of homogenised Johnson $B$ and $V$ magnitudes.
For the \object{Pleiades}, following \cite{Stauffer_etal:1998} we adopt an age of 120\,Myr.

\cite{Meibom_etal:2009} provide rotational periods and {\em UBVRI} photometry for 310 members of \object{M35}.
The age of M35 has been estimated to 150\,Myr \citep{vonHippel_etal:2002}, 175\,Myr \citep{Barrado_etal:2001}, and 180\,Myr \citep{Kalirai_etal:2003}.
Following \cite{Meibom_etal:2009}, we adopt an age of 150\,Myr.

Rotational periods and reddening corrected $(B-V)_0$ for \object{M34} are taken from \cite{Meibom_etal:2011}. The sample contains periods for 118 stars.
From the rotational period of slow-rotator sequence stars, adopting the Skumanich law, \cite{Meibom_etal:2011} estimated an age of 240\,Myr, which we also adopt here. 

\cite{Hartman_etal:2009} provide a clean sample of rotational periods for 372 stars in \object{M37}. 
Photometry for this cluster is from \cite{Kalirai_etal:2001}.
Following \cite{Hartman_etal:2008} we adopt an age of $550$\, Myr.

The rotational period data set of \cite{Delorme_etal:2011} includes 56 members of the \object{Hyades} and 35 members of \object{Praesepe}.
They also provide $BV$ photometry collected from the literature.
Following the discussion in \cite{Delorme_etal:2011}, we adopt an age of 600\,Myr for both clusters.

\cite{CollierCameron_etal:2009} present a photometric survey of rotation rates in the \object{Coma Berenices} cluster containing data for 27 members with available $BV$ photometry.
Following their discussion, we adopt an age for Coma Berenices equal to the Hyades and Praesepe age (600\,Myr) within the uncertainties.

\cite{Meibom_etal:2011a} provide rotational periods for 71 
cluster members of \object{NGC6811}.
Photometry for this cluster is provided by \cite{Janes_etal:2013}.
Following \citet[][and references therein]{Meibom_etal:2011a} we adopt an age of 1 Gyr.

Rotational periods and photometry for \object{NGC6819} is taken from \cite{Meibom_etal:2015}, who derived a gyro age of 2.5 Gyr in agreement with the estimate by \cite{Jeffries_etal:2013}. 

It is worth noting that the cluster ages adopted here do not differ significantly from the isochrone ages adopted by \cite{Kovacs:2015}.
In the present work, isochrone and rotational ages are assumed to coincide.

\section{Empirical description of the slow-rotator sequence evolution}
\label{sec:empirical}

\subsection{Empirical models}
\label{sec:empirical_models}

The main assumption in \cite{Barnes:2003, Barnes:2007} is that rotation periods on the slow-rotator sequence can be represented empirically by a relationship of the form
\begin{equation}
P = g(t) f(B-V) ,
\label{eq:factorisation}
\end{equation}
where $t$ is the stellar age and the colour $B-V$ is taken as a proxy of the stellar mass.
Other authors fitted the function $P=P(t,B-V)$ to different data sets \citep[e.g.][]{Meibom_etal:2009,Mamajek_Hillenbrand:2008}, maintaining the same form and factorisation as in \cite{Barnes:2007}.
Both \cite{Barnes:2007} and \cite{Mamajek_Hillenbrand:2008} derived their color-period relationship on a few well studied clusters (e.g. $\alpha$ Persei, Pleiades, Hyades) while the fit by
\cite{Meibom_etal:2009} is based on a single cluster (M35).

\cite{Barnes:2003} proposed that the mass dependence of the rotation periods in open clusters can be simply attributed to the moments of inertia of either the whole star or of the outer convection zone. 
This was later found inadequate by \cite{Barnes_Kim:2010}, and led \cite{Barnes:2010} to the formulation of a different non-linear empirical model, which provides the gyro-age of a star as
\begin{equation}
t = \frac{\tau}{k_C} \ln\left(\frac{P}{P_0}\right) + \frac{k_I}{2 \tau} \left(P^2 - P_0^2 \right),
\label{eq:Barnes2010}
\end{equation} 
where $k_C$ and $k_I$ are two dimensionless constants, constrained by the observations, and $P_0$ is the initial rotation period, taken as the period on the zero-age main sequence.
In the slow-rotator sequence limit we are concerned with, this semi-empirical formula gives
\begin{equation}
P = \sqrt{P_0^2 + \frac{2 \tau t}{k_I}} \,,
\label{eq:I-limit}
\end{equation}
which predicts that, asymptotically, $P \rightarrow g(t) f(B-V)$ with $g(t) = \sqrt{t}$ and $f(B-V) = \sqrt{2 \tau / k_I}$.
The \cite{Barnes:2010} model gives a description of the mass-dependent evolution of fast-rotating stars to the slow-rotator sequence. 
An alternative description is the metastable dynamo model (MDM) of \cite{Brown:2014}, which assumes a stochastic transition between two different dynamo regimes and is rooted in an original idea of \cite{Barnes:2003}.

\begin{figure*}
\begin{center}
\includegraphics[width=0.44\textwidth]{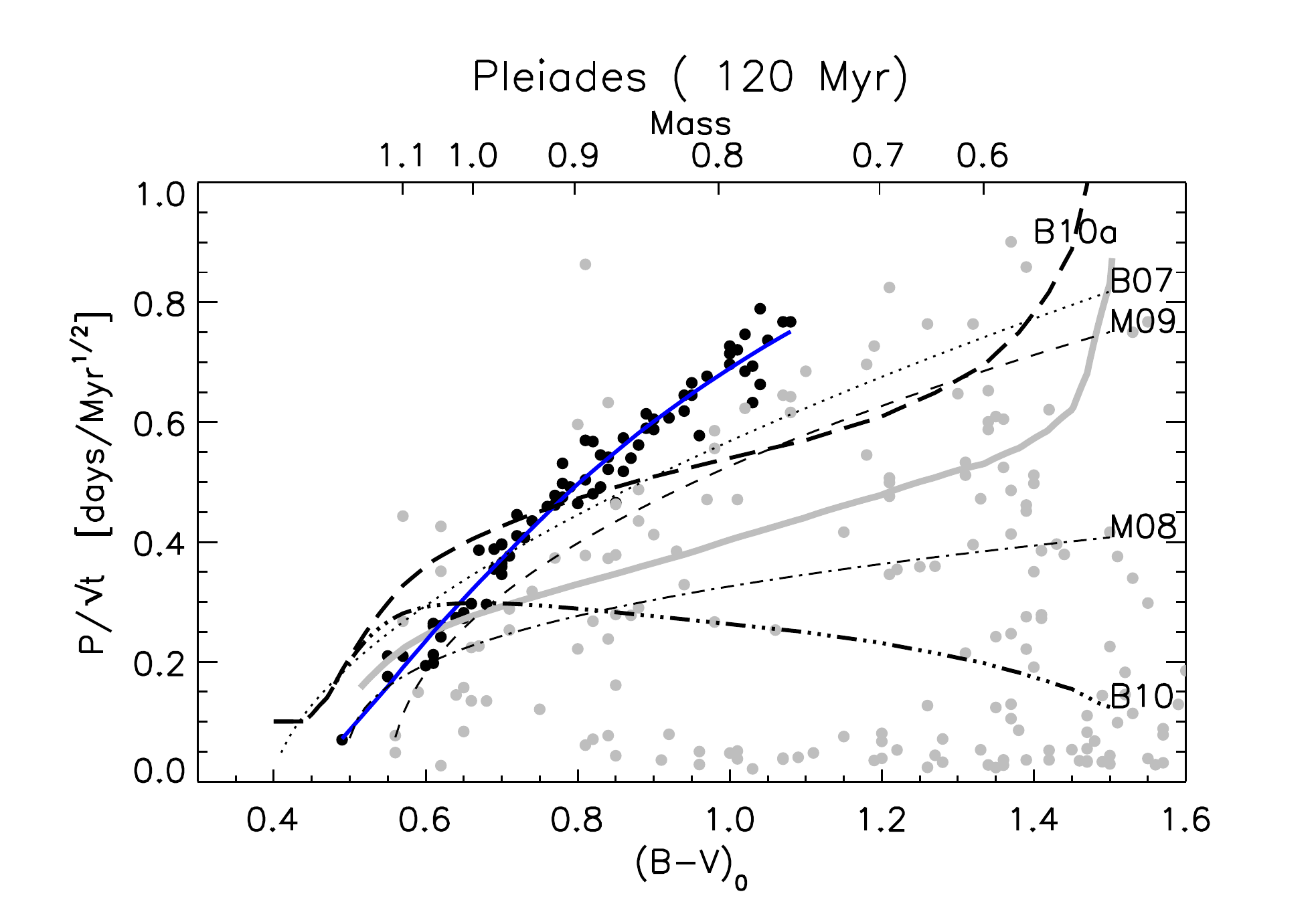}
\includegraphics[width=0.44\textwidth]{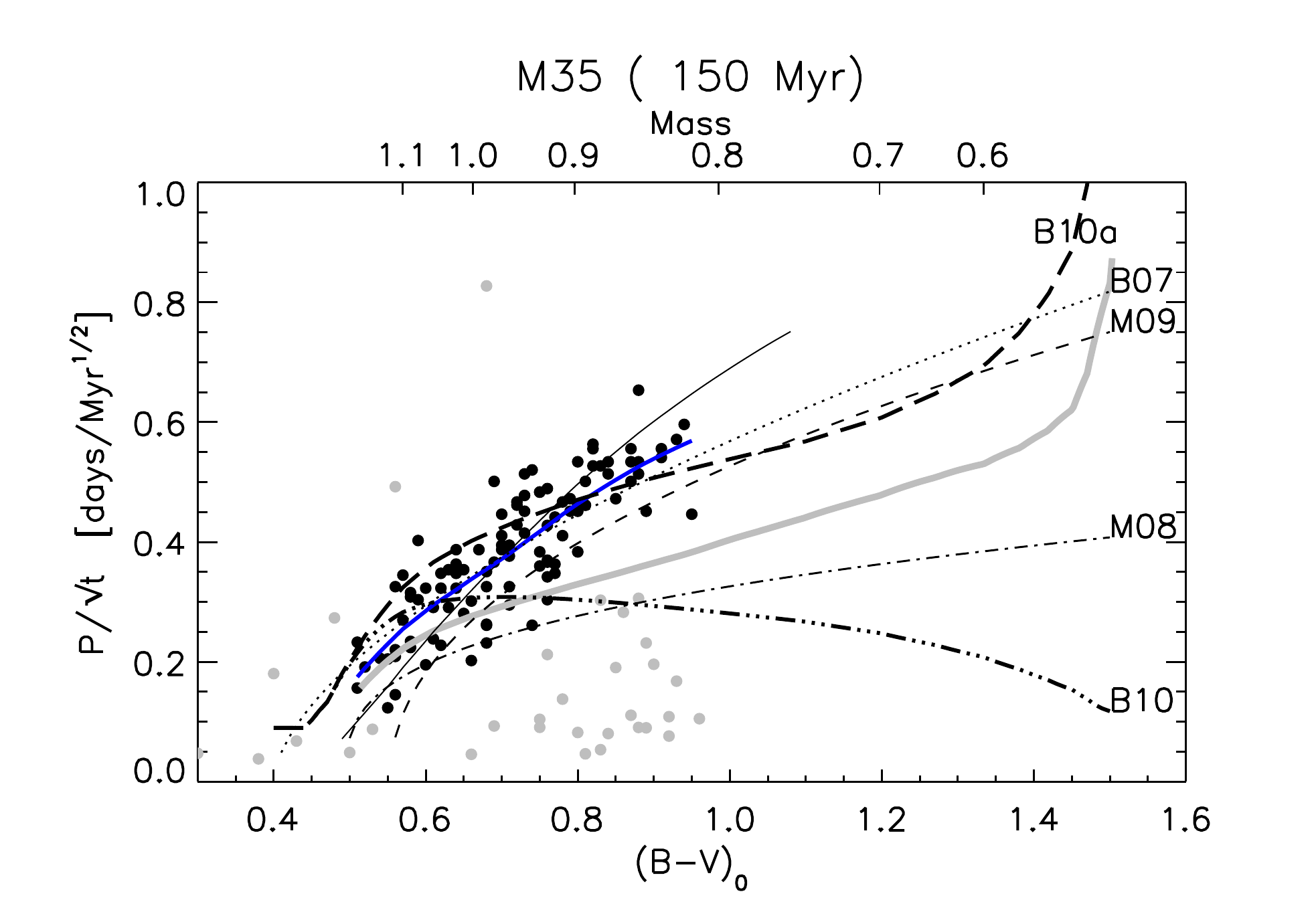}
\includegraphics[width=0.44\textwidth]{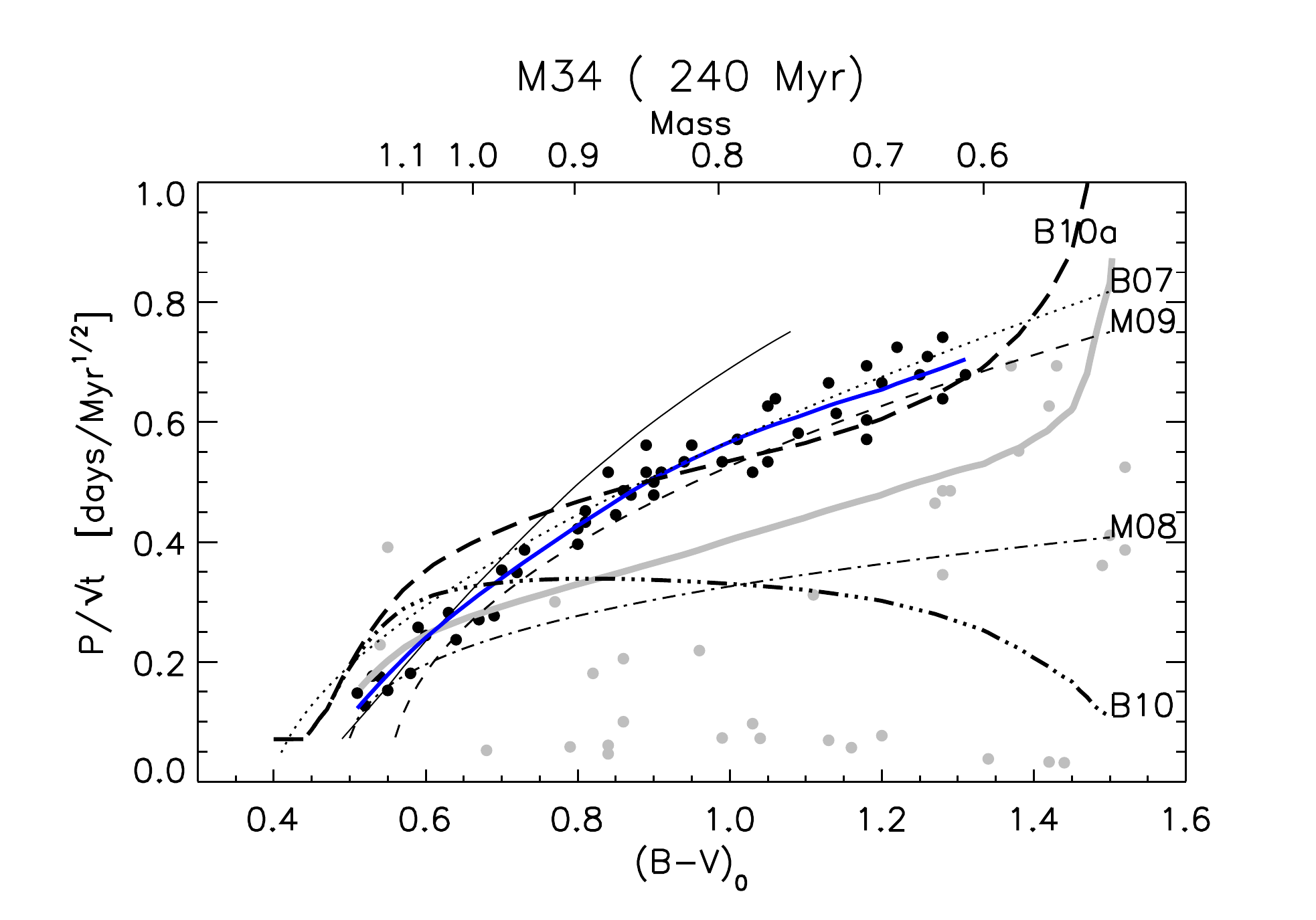}
\includegraphics[width=0.44\textwidth]{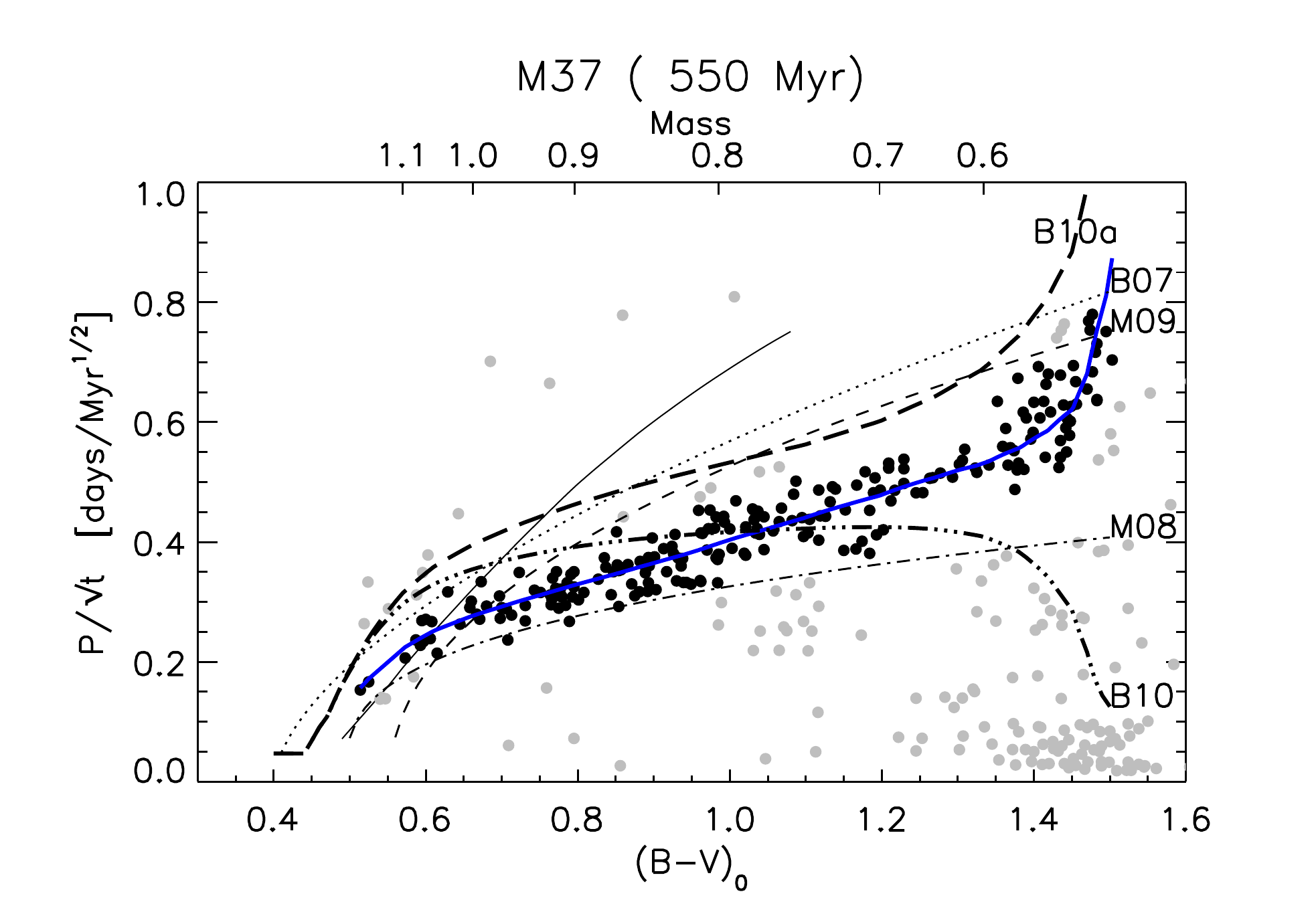}
\includegraphics[width=0.44\textwidth]{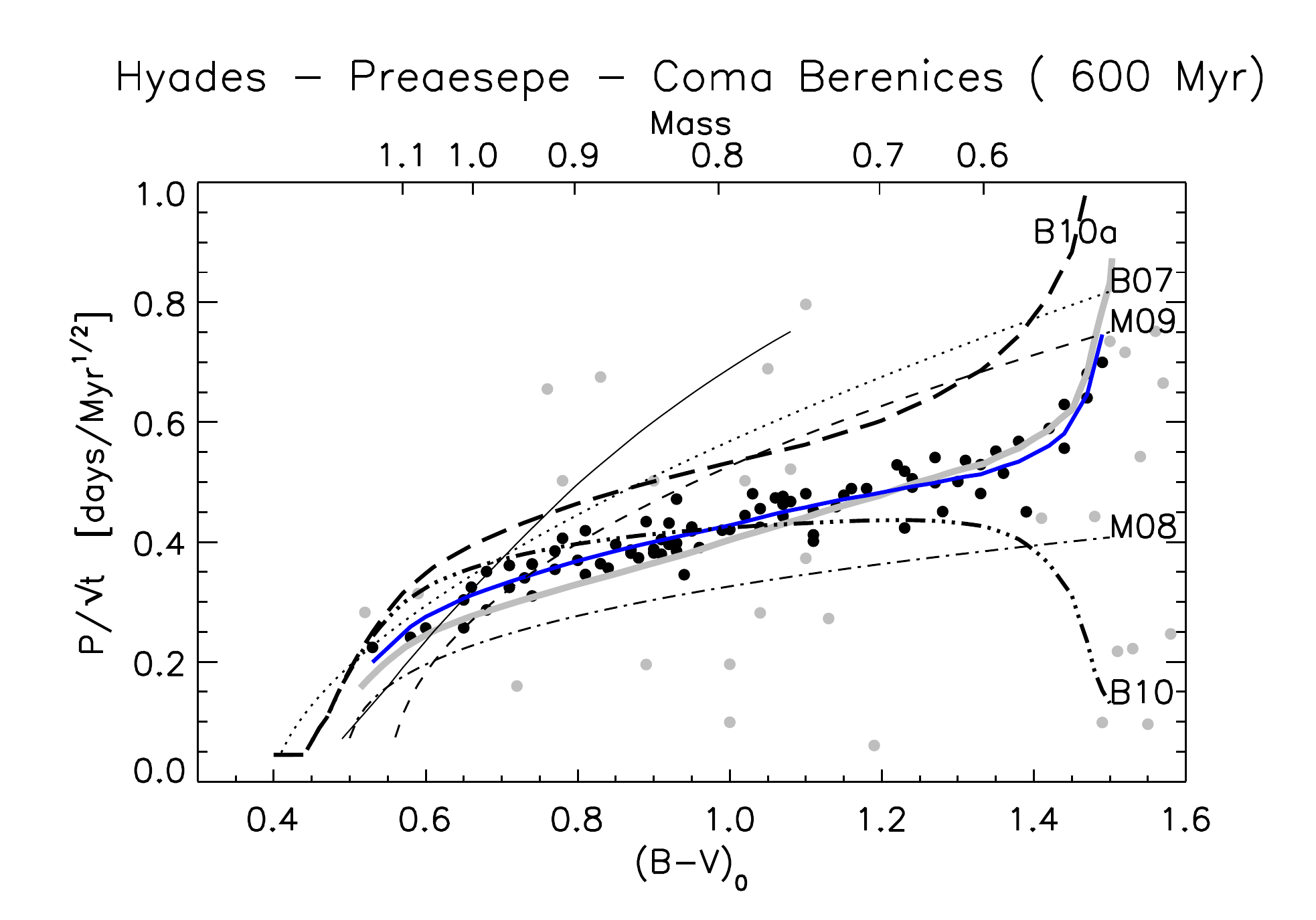}
\includegraphics[width=0.44\textwidth]{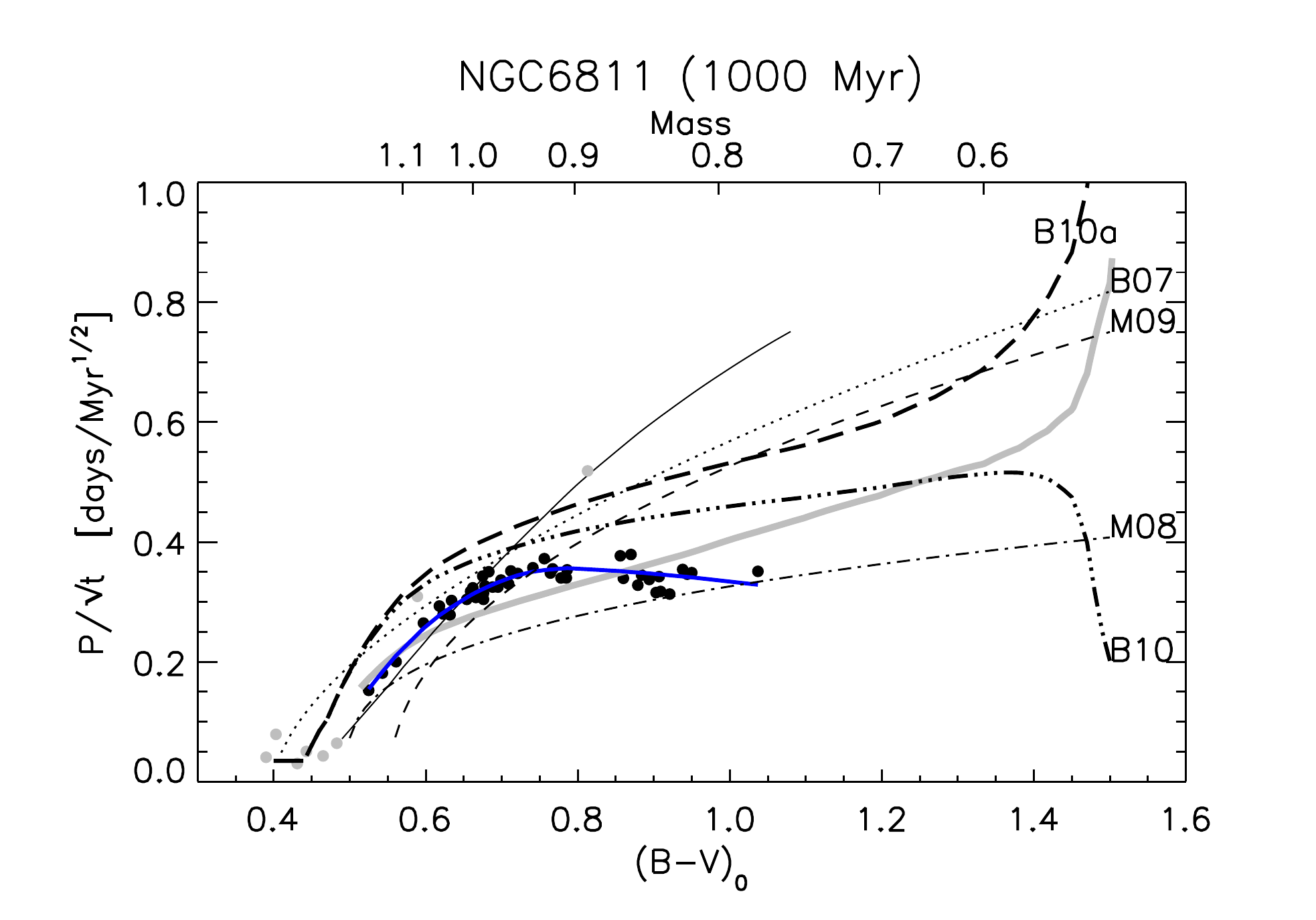}
\includegraphics[width=0.44\textwidth]{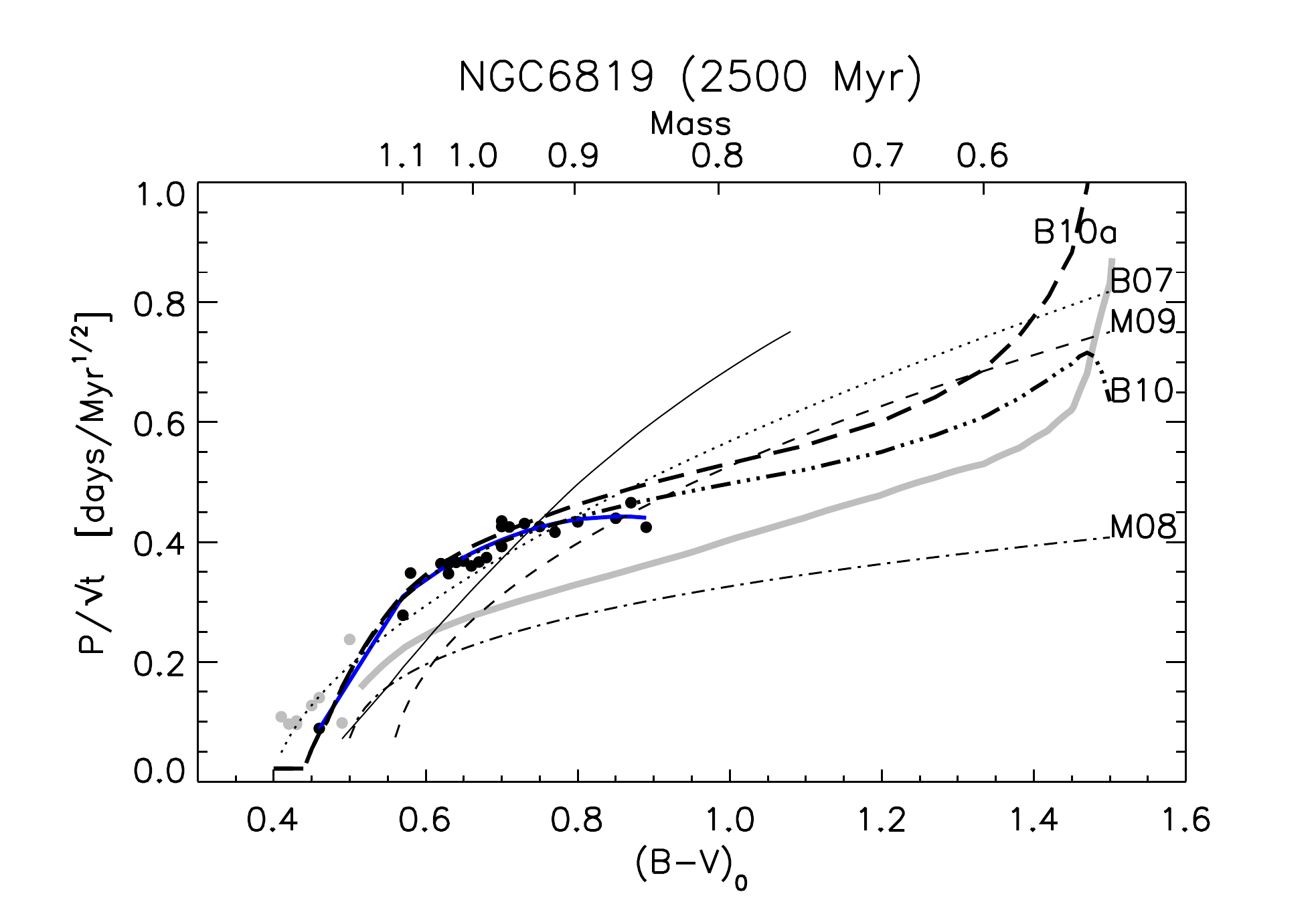}
\caption{
Rotation period scaled to the square root of age (in Myr) vs. $(B-V)$. 
Periods of stars defining the slow-rotator sequence are represented with black dots, all others with grey dots. 
Our non-parametric fit of the slow-rotator sequence is shown as blue thick solid lines, with the Pleiades and M37 fits repeated in all panels as reference (black thin solid line and grey thick solid line respectively). 
The dotted line represent the $f(B-V)$ function of \cite{Barnes:2007} (B07), the dashed line that of \cite{Meibom_etal:2009} (M09), the dot-dashed that of \cite{Mamajek_Hillenbrand:2008} (M08), the dash-triple-dotted line the inverted \cite{Barnes:2010} relationship (Eq.\,\ref{eq:Barnes2010}, B10), the long-dashed line the \cite{Barnes:2010} asymptotic slow-rotator sequence limit (Eq.\,\ref{eq:I-limit}, B10a).
}
\label{fig:GyA_periods}
\end{center}
\end{figure*}

The color-period diagrams for the clusters studied here are shown in Fig.\,\ref{fig:GyA_periods}.
The stellar mass is estimated using the color-mass relationship of \cite{Barnes_Kim:2010},
based on the effective temperature-color transformation of \cite{Lejeune_etal:1997,Lejeune_etal:1998}.
Periods are scaled to the square root of age (in Myr), so that data points would overlap in diagrams at different ages if the rotational evolution were exactly Skumanich-type.
The slow-rotator sequence is easily identifiable in all clusters, with the only exception of the Pleiades and M35, for which some ambiguity may arise (see Sect.\,\ref{sec:non-parametric}).
From the comparison of the observed slow-rotator sequence at different ages it is evident that:
\begin{itemize} 
\item Stars with $M < 1.0\, M_{\odot}$ slow down at a slower rate than predicted by the Skumanich law until $t \sim 0.55$ Gyr;
\item From $0.55$ to $2.5$ Gyr, stars slow down at a faster rate than predicted by the Skumanich law, except for the NGC6811 stars in the $0.8 \le M/M_{\odot} \le 0.9$ mass range;
\item There is no evidence of stars of mass $M < 0.8\, M_{\odot}$ on the slow-rotator sequence at ages younger than $0.24$ Gyr.
\end{itemize}

The current observed evolution of the slow-rotator sequence from $0.1$ to $2.5$ Gyr shows that the factorization expressed by Eq.\,(\ref{eq:factorisation}) does not hold, at least before $0.55$ Myr, because such a relationship would imply that the shape of the slow-rotator sequence is maintained throughout its evolution.

The data available contains periods for stars on the slow-rotator sequence with mass as low as $M \approx 0.5 M_{\odot}$ for ages $\approx$0.6 Gyr, but no data for stars with $M < 0.8 M_{\odot}$ is available at later ages.

In Fig.\,\ref{fig:GyA_periods} we also report the $f(B-V)$ empirical functions obtained by \cite{Barnes:2007}, \cite{Meibom_etal:2009}, and \cite{Mamajek_Hillenbrand:2008}, as well as the inversion of Eq.\,(\ref{eq:Barnes2010}) from \cite{Barnes:2010} and its asymptotic limit in Eq.\,(\ref{eq:I-limit}).
The comparison with observations shows that such relationships can describe the slow-rotator sequence only on limited age and mass ranges.

With the exception of stars of $0.8 \le M \le 0.9\, M_{\odot}$ in NGC6811, the slow-rotator sequence lies above the $\approx 0.55$ Gyr sequence in the $P/\sqrt{t}$ vs. $(B-V)$ diagram.
In the framework of the two-zone model (Sect.\,\ref{sec:tzm}), this behavior could be attributed to the transport of angular momentum from the stellar core to the envelope in the earlier evolution after the ZAMS.
We shall verify this in Sect.\,\ref{sec:tzm_fitting}. 

\subsection{Non-parametric empirical fitting}
\label{sec:non-parametric}

\begin{table*}
   \centering
   \begin{tiny}
      \caption{Rotational periods from the non-parametric fit. $\sigma$ is the standard deviation, $p$ the Pearson test probability that the residual distribution is normal.}
      \label{tab:periods}
      \begin{tabular}{lrrrrrrrrrrrrrr}
      \hline
cluster       &  age   &   \multicolumn{11}{c}{$M/M_{\odot}$}                                                          & $\sigma$ & $p$ \\
	                    &       & 1.10 &  1.05 &  1.00 & 0.95  & 0.90  &  0.85 &  0.80 &  0.75 &  0.70 & 0.65  & 0.60  &      &     \\
\hline
           Pleiades &  120 &  2.09 &  2.75 &  3.53 &  4.42 &  5.39 &  6.41 &  7.42 & \dots & \dots & \dots & \dots &  0.35 &   0.63 \\
                M35 &  150 &  3.12 &  3.62 &  4.16 &  4.82 &  5.64 &  6.46 & \dots & \dots & \dots & \dots & \dots &  0.76 &   0.80 \\
                M34 &  240 &  3.18 &  3.92 &  4.72 &  5.60 &  6.57 &  7.65 &  8.65 &  9.48 & 10.11 & 10.62 & \dots &  0.53 &   0.82 \\
                M37 &  550 &  5.21 &  5.91 &  6.50 &  7.08 &  7.70 &  8.42 &  9.32 & 10.32 & 11.19 & 11.91 & 12.43 &  0.94 &   0.68 \\
             Hyades &  600 & \dots & \dots &  7.62 &  8.35 &  9.07 &  9.82 & 10.64 & 11.57 & 12.26 & 12.72 & 13.04 &  0.51 &   0.52 \\
           Praesepe &  600 & \dots &  6.53 &  7.36 &  8.16 &  8.90 &  9.55 & 10.02 & 10.48 & 11.09 & 11.77 & \dots &  0.63 &   0.89 \\
      ComaBerenices &  600 &  6.68 &  7.49 &  8.12 &  8.70 &  9.29 & 10.07 & 10.75 & 11.07 & 11.23 & 11.37 & \dots &  0.52 &   0.26 \\
Hydes-Praesepe-Coma &  600 &  6.12 &  6.95 &  7.65 &  8.31 &  8.99 &  9.68 & 10.38 & 11.20 & 11.77 & 12.22 & 12.58 &  0.73 &   0.68 \\
            NGC6811 & 1000 &  7.07 &  8.58 &  9.90 & 10.93 & 11.30 & 11.03 & 10.64 & \dots & \dots & \dots & \dots &  0.46 &   0.45 \\
            NGC6819 & 2500 & 15.46 & 17.41 & 19.12 & 20.75 & 21.87 & 22.07 & \dots & \dots & \dots & \dots & \dots &  0.88 &   0.92 \\
\hline
      \end{tabular}
	\end{tiny}
\end{table*}

\begin{figure*}
\begin{center}
\includegraphics[width=50mm]{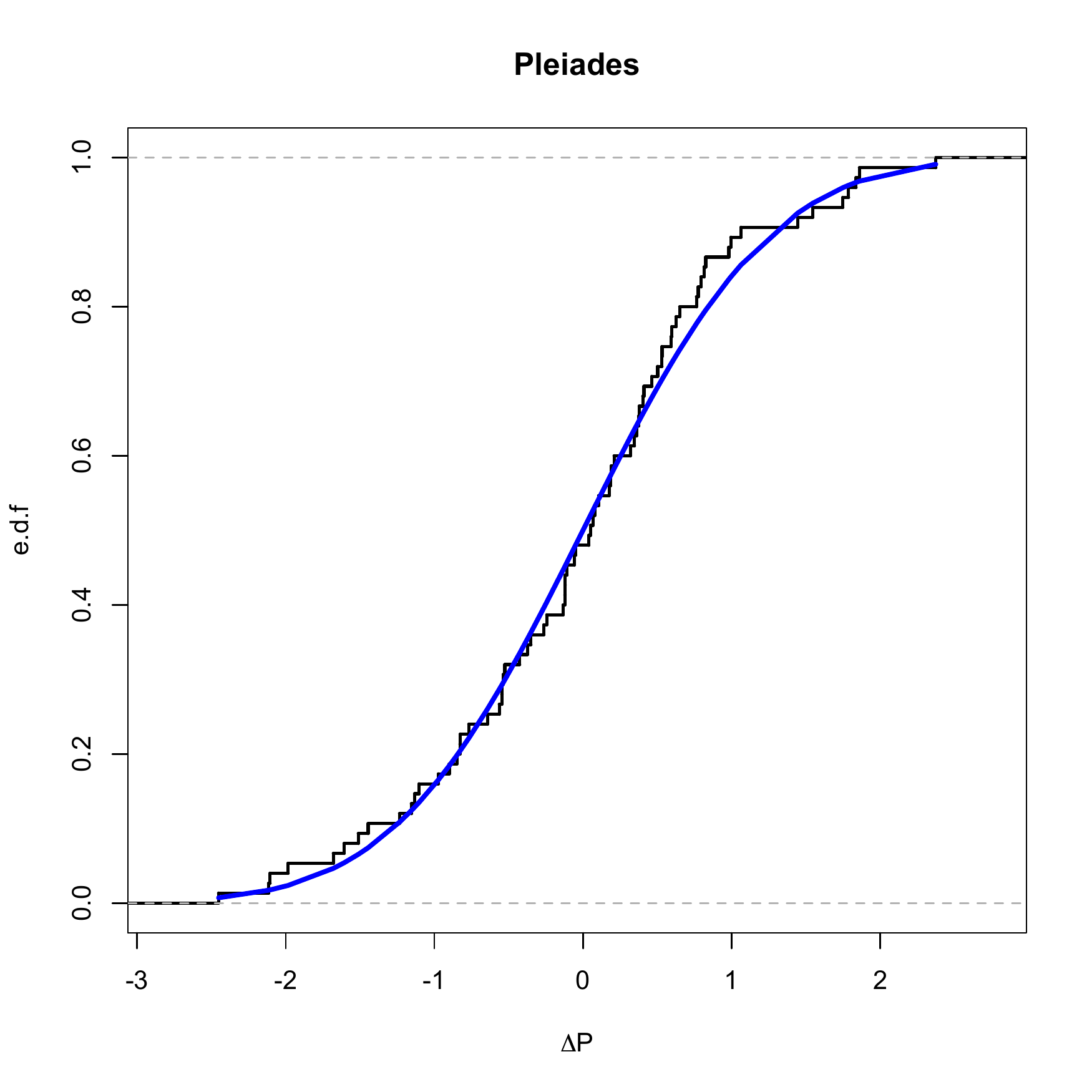}
\includegraphics[width=50mm]{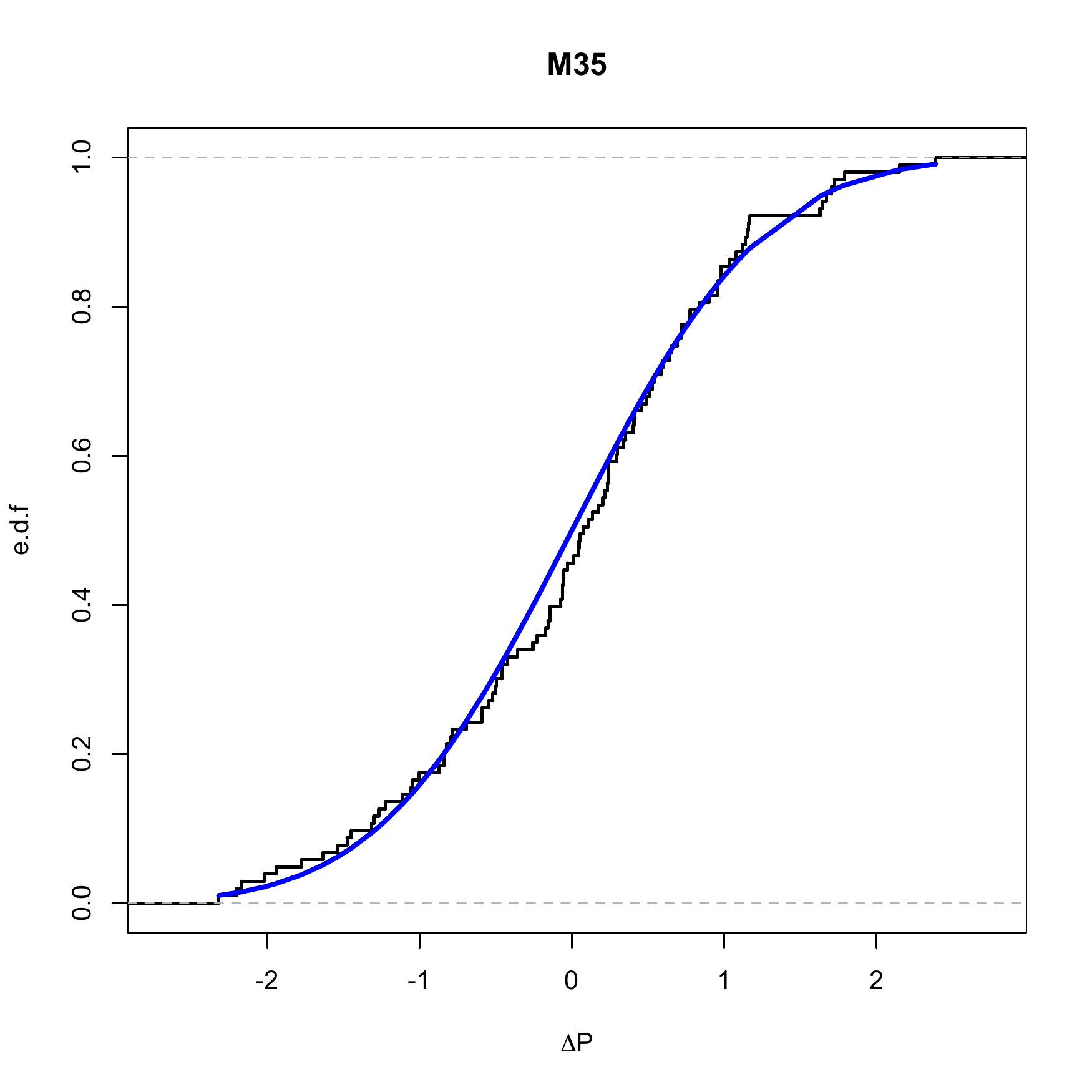}
\includegraphics[width=50mm]{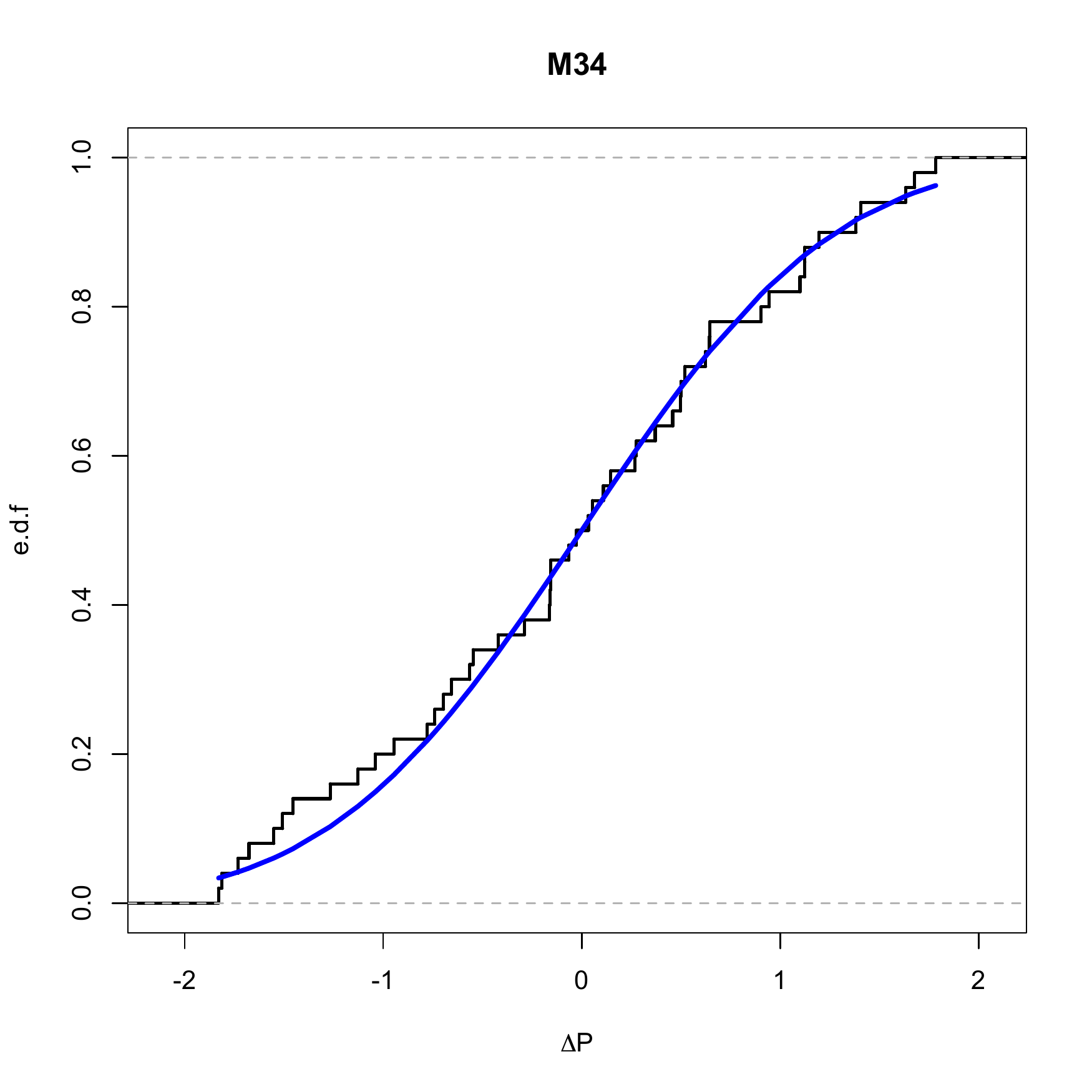}
\includegraphics[width=50mm]{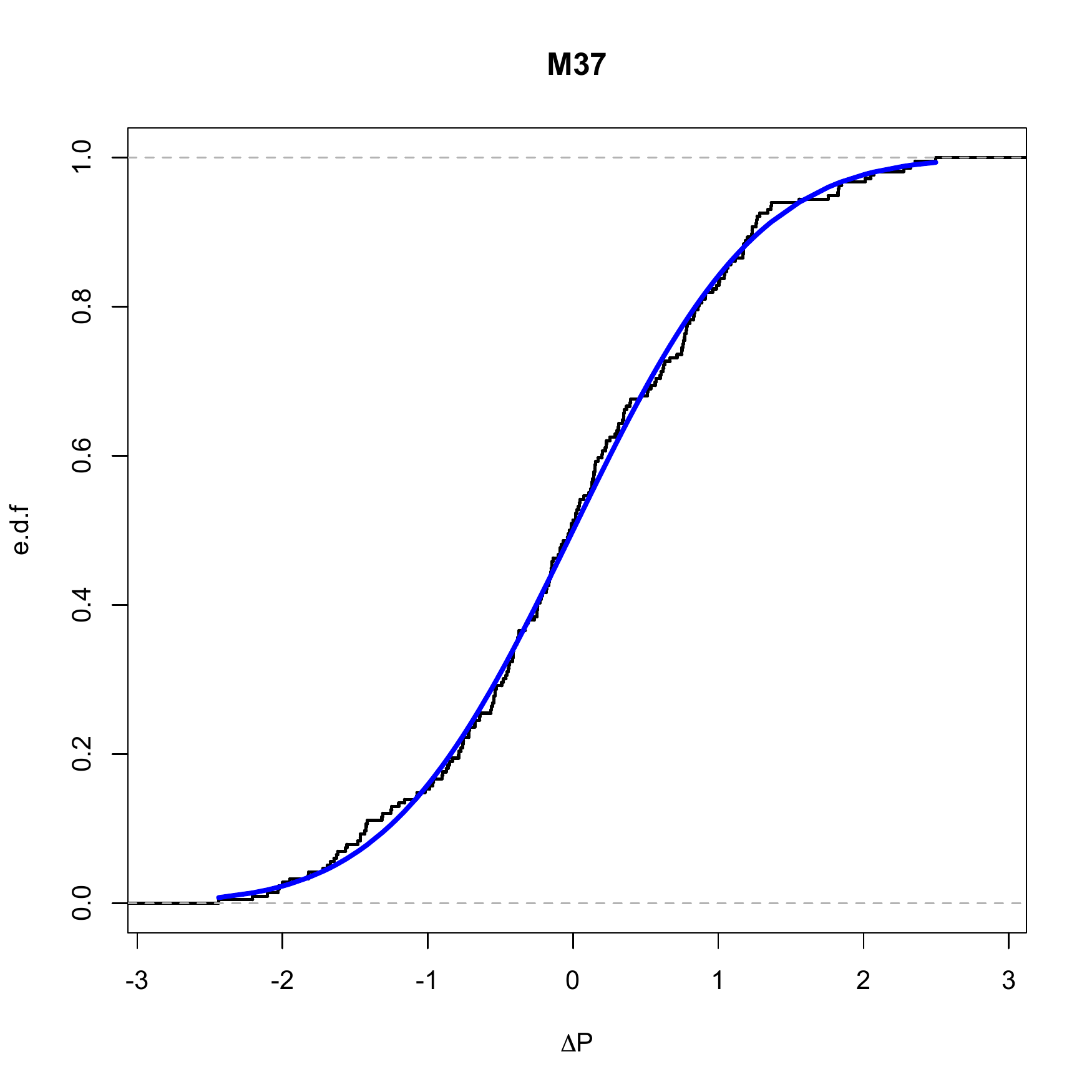}
\includegraphics[width=50mm]{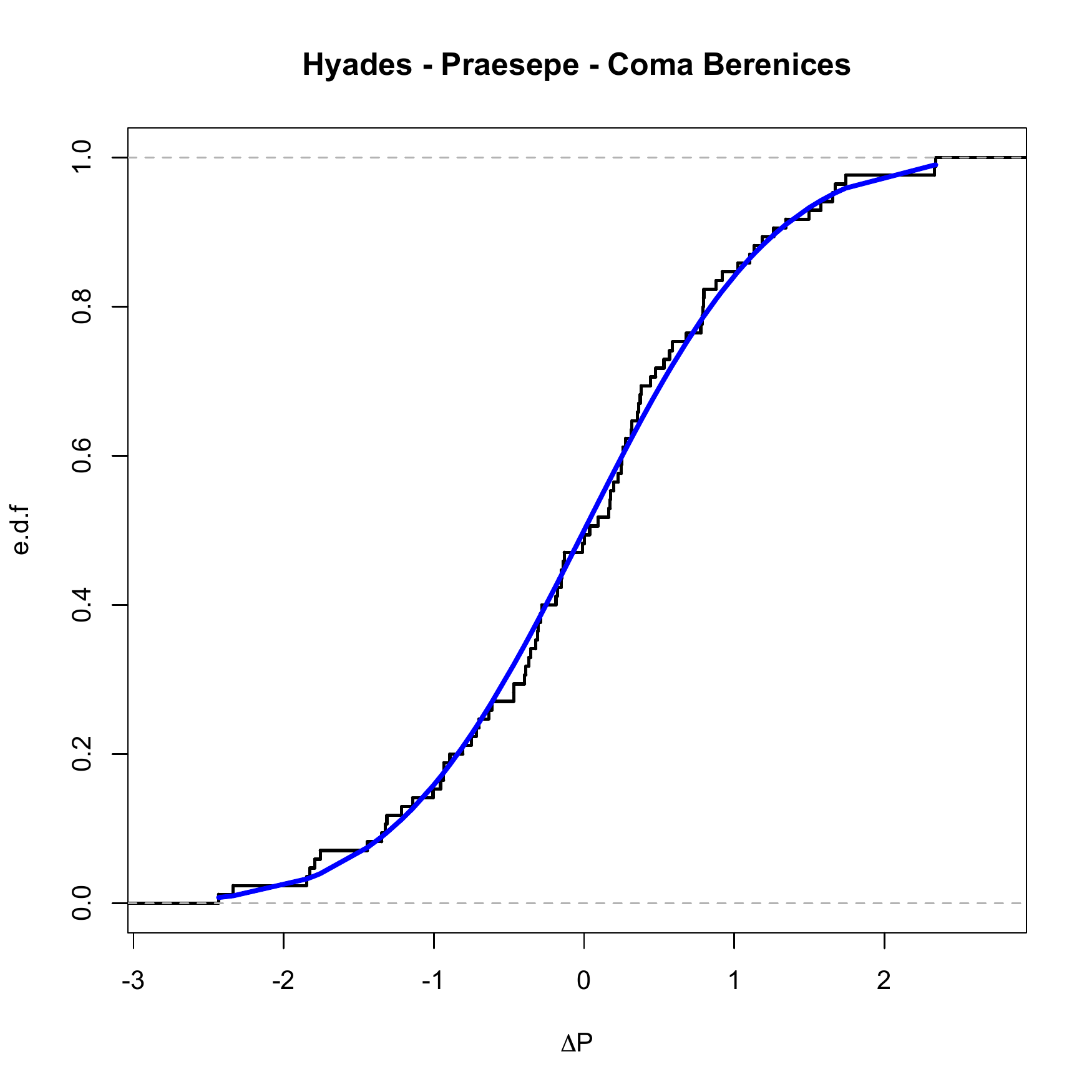}
\includegraphics[width=50mm]{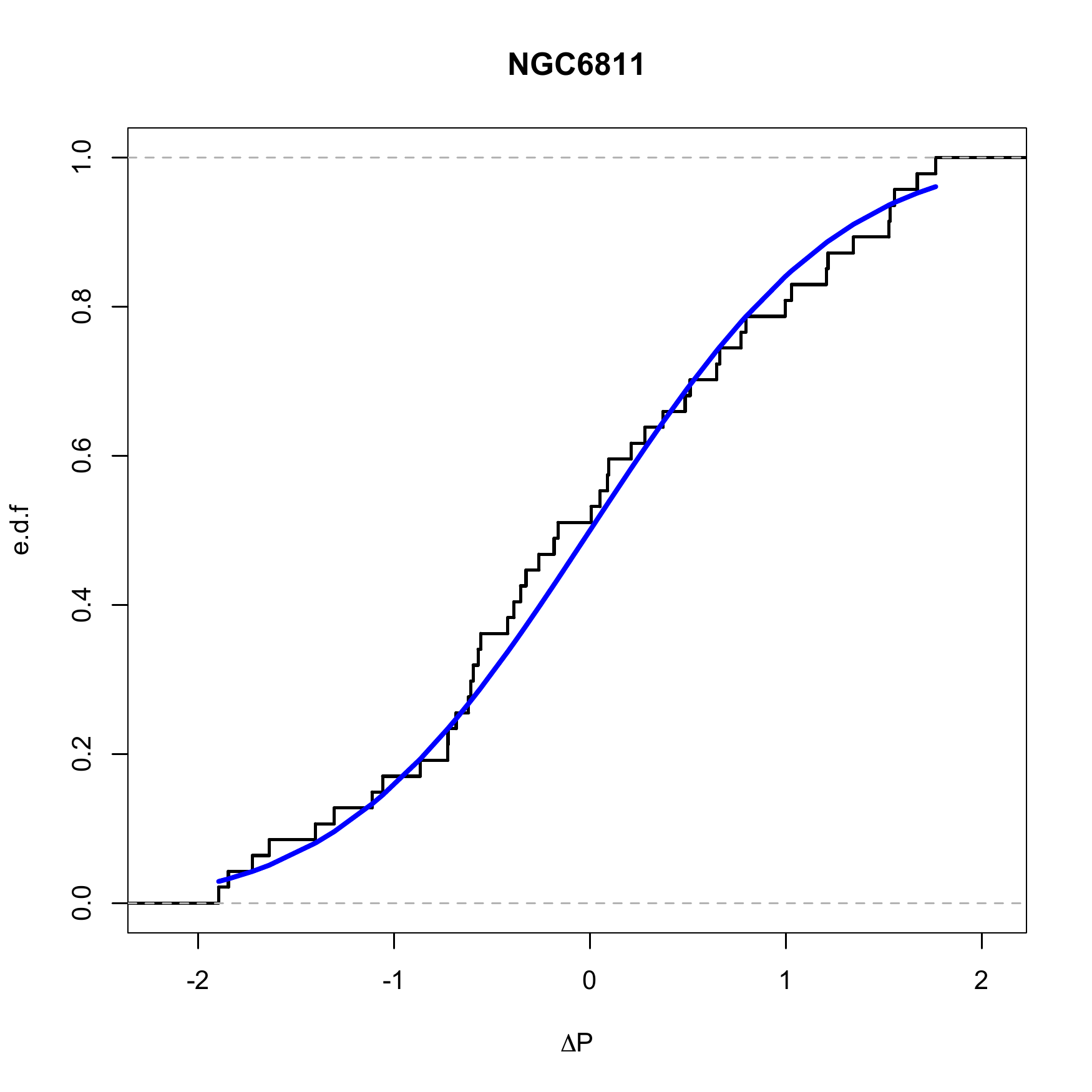}
\includegraphics[width=50mm]{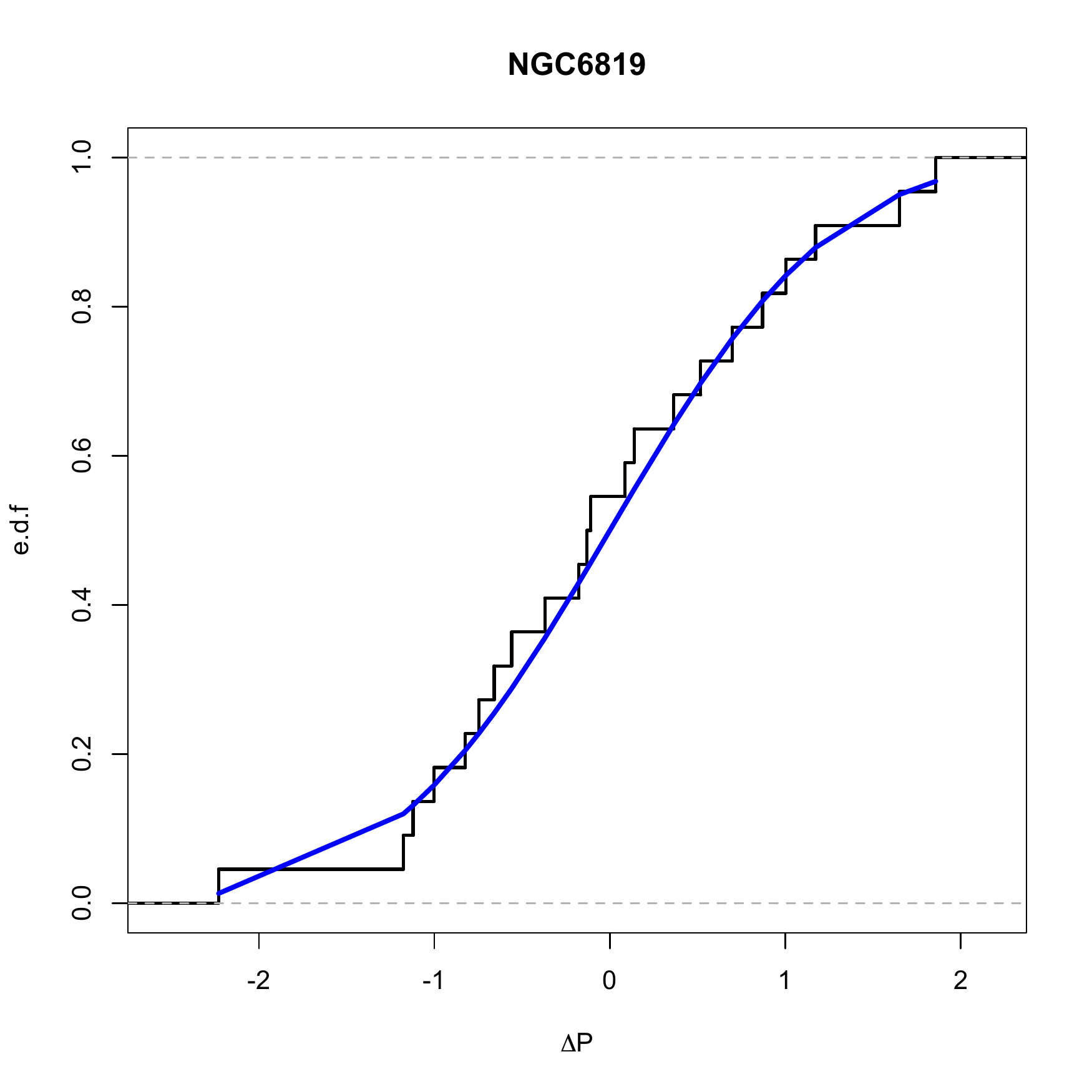}
\caption{Comparison of the empirical distribution function of the non-parametric fit residuals with the normal distribution function. Residuals are normalized to the standard deviation.}
\label{fig:GyA_edf}
\end{center}
\end{figure*}

In order to overcome the limitations of the empirical modeling discussed in the Sect.\,\ref{sec:empirical_models}, we perform a non-parametric fit to the slow-rotator sequence.
To this end, we need a criterion for selecting the stars belonging to this sequence. 
For the older clusters, the selection is quite simple as only some outliers and a few remaining fast-rotators need to be excluded.
For the younger clusters, on the other hand, the slow-rotator sequence over-density is still easily identifiable in the color-period diagram, but the separation with stars approaching the slow-rotator sequence is somewhat arbitrary and prone to subjective choices.
Slow-rotator sequence membership may become even more ill-defined if, following \cite{Barnes:2010}, we consider the sequence as an asymptotic limit.

We note, however, that non-parametric fits to the older slow-rotator sequences produce residuals whose distribution can be approximated with a normal distribution.
To trace back the sequence to the younger clusters with a criterion as homogeneous as possible, we set the width of the slow-rotator sequence as the maximum width that produces a distribution of residuals with a normality probability (see below) of at least 30\% (see Fig.\,\ref{fig:GyA_edf} and Table\,\ref{tab:periods}).
We use the local polynomial regression fit \citep[LOESS,][]{Cleveland_etal:1992} with smoothing parameter between $0.8$ and $1.0$, and a Pearson normality test \citep{Moore:1986,Thode:2002}, as implemented in the statistical package \texttt{R}.
In order to improve the fitting at the higher and lower end of the sequence, where the curvature is higher and a non-parametric fit would require a much denser data set, we use Eq.\,(\ref{eq:I-limit}) as a normalization  function, adjusting $k_I$ in order to mimic the slow-rotator sequence at $\approx 0.6$\,Gyr.

Our non-parametric fits are shown in Fig.\,\ref{fig:GyA_periods}, where the stars selected as members of the slow-rotator sequence are also highlighted.
The non-parametric fit is used to derive periods for a grid of stellar masses, reported in Table\,\ref{tab:periods} together with the standard deviation and the normality test probability.
The comparisons of the empirical distribution functions of the residuals with normal distributions are shown in Fig.\,\ref{fig:GyA_edf}.

Our selection of stars belonging to the slow-rotator sequence in younger clusters is more restrictive than that adopted, for instance, by \cite{Barnes:2010}.
It should be considered, however, that the aim of the present work is to study in detail the rotational evolution of the slow-rotator sequence, under the assumption that this represents some equilibrium state or asymptotic behavior, deliberately ignoring fast rotators or stars still approaching the slow-rotator sequence.
We also ignore periods much larger than the periods of the slow-rotator sequence, assuming that they are either incorrect or that these stars are not cluster members.
In other words, we assume that the peak of the over-density of slow rotators in the color-period diagram corresponds to the maximum probability of having reached the equilibrium/asymptotic state and that stars in or near such a state have periods distributed normally around the peak.
Such an approximation has the advantage of allowing a characterization of the width of the sequence using the standard deviation, which can then be used as input in our parametric fit described in Sect.\,\ref{sec:tzm_fitting}.
It turns out that such an approximation describes remarkably well the distribution in the older clusters, and that, by an appropriate selection, it can be extended to the younger clusters as well.
For our purposes, it can safely be assumed that the standard deviation of the sequence includes both observational uncertainties and the intrinsic physical width due to the ongoing convergence onto the slow-rotator sequence.

\section{Two-zone models Monte Carlo Markov-chain fitting to the data}
\label{sec:tzm_fitting}

\subsection{Two-zone rotational evolution models}
\label{sec:tzm}

In our modeling of the rotational evolution of solar-like stars (whose structure is characterized by an inner radiative region, surrounded by a convective envelope), we adopt the theoretical framework of two-zone models \citep[TZMs; see, e.g.][]{MacGregor_Brenner:1991,Keppens_etal:1995, Allain_etal:1998,Spada_etal:2011}.
The main assumption of the TZM is that the radiative interior and the convective envelope of the star rotate as rigid bodies, with angular velocities $\Omega_c$ and $\Omega_e$, respectively; the rotational state of the whole star at a given time is thus completely specified by these two quantities.

During the pre-main sequence, the rotational evolution is dominated by evolutionary changes to the stellar structure (i.e., overall contraction, appearance and growth of a radiative core), and the interaction with the circumstellar disc. 
From the zero age main sequence onwards, the interior structure changes very little, and the rotational evolution essentially depends on the interplay between the angular momentum loss at the surface, due to the magnetized stellar wind, and the angular momentum exchange between core and envelope.
These main physical ingredients are included in our models as follows:

\begin{itemize}
\item The radiative core grows at the expenses of the surrounding envelope, thus subtracting from it the angular momentum \citep{Allain_etal:1998}: $\frac{2}{3}\dot M_c R_c^2 \Omega_e$, where $M_c$ and $R_c$ are the mass and the radius of the core, respectively (in the following, we adopt the dot notation for derivatives with respect to time);
\item The star-disc interaction is assumed to be very efficient, capable of keeping the surface angular velocity constant during the entire disc lifetime, $\tau_{\rm disc}$ \citep[the so-called disc-locking approximation, see][]{Koenigl:1991};
\item The rate of angular momentum loss from the surface due to the magnetized stellar wind is specified by the wind braking law $\dot J_{\rm wb}$ (see Section~\ref{secwb} for details);
\item Over the coupling timescale $\tau_{\rm cp}$, the amount of angular momentum $\Delta J = \dfrac{I_cI_e}{I_c+I_e} (\Omega_c-\Omega_e)$ is transferred from the radiative interior to the convective envelope \citep[see][]{MacGregor_Brenner:1991}.
\end{itemize}
The TZM equations for $\Omega_c$ and $\Omega_e$ are therefore: 
\begin{itemize}
\item For $t \leq \tau_{\rm disc}$:
\begin{equation}
\left\{
\label{eqtzmdisc}
\begin{array}{rl}
I_c\dot \Omega_c &= + {2}/{3} \dot M_c R_c^2 \, \Omega_e - \dot I_c \Omega_c; 
\\
\dot \Omega_e &= 0;   
\end{array}
\right.
\end{equation}
\item For $t > \tau_{\rm disc}$:
\begin{equation}
\left\{
\label{eqtzm}
\begin{array}{rll}
 I_c \dot \Omega_c = & + {2}/{3} \dot M_c R_c^2 \,\Omega_e - {\Delta J}/{\tau_{\rm cp}}  - \dot I_c \Omega_c; & 
 \\
  I_e \dot \Omega_e = & - {2}/{3} \dot M_c R_c^2 \, \Omega_e + {\Delta J}/{\tau_{\rm cp}}   - \dot I_e \Omega_e  & + \; \dot J_{\rm wb}  .  
\end{array}
\right.
\end{equation}
\end{itemize}
In the equations above, we have introduced the moments of inertia, $I_c$ and $I_e$, of the radiative zone and of the convective envelope, respectively.
Note the appearance of terms involving the time derivatives of the moments of inertia ($\dot I_e \Omega_e $, etc.), as required by angular momentum conservation when significant structural changes take place.

We begin the integration of equations \eqref{eqtzmdisc} and \eqref{eqtzm} when the star is still fully convective, when it is reasonable to assume that the star is rotating as a solid body. 
We thus set the initial conditions in terms of a single parameter, the initial period $P_0$:
\begin{eqnarray*}
\Omega_c(t_0) = \Omega_e(t_0) = \frac{2\pi}{P_0}.
\end{eqnarray*}
We extract the evolution of stellar structure quantities, 
$R_c$, $M_c$, etc., 
from stellar models of appropriate mass, calculated using the Yale Rotational stellar Evolution Code (YREC), in its non-rotational configuration \citep{Demarque_etal:2008}.
The stellar models were calculated with initial composition and mixing length parameter $\alpha_{\rm MLT}$ calibrated on the Sun.
Assuming the \citet{Grevesse_Sauval:1998} value of the solar metallicity, $(Z/X)_{\odot,\rm GS98}\approx 0.0230$, our solar calibration gives $X_0=0.7039$, $Y_0=0.2773$, $Z_0=0.0188$, $\alpha_{\rm MLT}=1.826$.

\subsection{Wind braking laws}
\label{secwb}

In our TZM framework, the wind braking law prescribes the dependence of $\dot J_{\rm wb}$ on the stellar mass (either directly or indirectly, i.e., through a dependence on other stellar structure variables), and on the surface angular velocity $\Omega_e$.
In this work, we focus on the following wind braking laws:
\begin{enumerate}
\item 
The \citet{Kawaler:1988} law as modified by \citet{Chaboyer_etal:1995}:
\begin{equation}
\label{kawalerwb}
\dot J_{\rm wb} =  
\left\{
\begin{array}{cc}
- K_w \left(\dfrac{R_*/R_\odot}{M_*/M_\odot}\right)^{1/2} \Omega_e^3 & \mathrm{for} \; \Omega_e < \Omega_{\rm sat} \\
\\
- K_w \left(\dfrac{R_*/R_\odot}{M_*/M_\odot}\right)^{1/2} \Omega_{\rm sat}^2\Omega_e & \mathrm{for} \; \Omega_e \ge \Omega_{\rm sat} \\
\end{array}
\right. .
\end{equation}
Note that the dependence on $\Omega_e^3$ reproduces the empirical Skumanich law if the star rotates as a solid body.
Our scaling of $K_w$ requires that the numerical value given in Sect.\,\ref{sec:results} must be multiplied by $1.11 \cdot 10^{47}$\,g\,cm$^2$\,s to convert it into cgs units.
\item 
A hybrid braking law, which combines the classical $\Omega_e$ dependence of \citet{Kawaler:1988} with the mass dependence suggested by \cite{Barnes_Kim:2010} and \cite{Barnes:2010} for the slow-rotator sequence:
\begin{equation}
\label{modkawwb}
\dot J_{\rm wb} =  
- K_w \left(\frac{I_* \tau}{{I_{\odot} \tau_{\odot}}} \right) \Omega_e^3 
\end{equation}
where $I_*$ and $I_{\odot}$ are the moment of inertia of the whole star and of the Sun, respectively, and $K_w$ is an adjustable parameter with the same scaling as above.
\item 
The braking law proposed by \citet{Gallet_Bouvier:2013}. 
Their starting point is the very general expression 
$\dot J_{\rm wb} \propto \Omega_e \dot M_{\rm wind}\, r_A^2$, 
where $\dot M_{\rm wind}$ is the mass loss due to the stellar wind and $r_A$ is the Alfv\`en radius, i.e., the radius of the (approximately) spherical region around the star where the wind is dominated by the magnetic field \citep{Weber_Davis:1967}.
\citet{Gallet_Bouvier:2013} use an expression derived from the numerical simulation of \citet{Matt_etal:2012a} to eliminate the Alfv\`en radius from the braking law; their final result is \citep[see Sect.\,3.3 of][for details]{Gallet_Bouvier:2013}:
\begin{equation}
\label{galletwb}
\dot J_{\rm wb} = K_1^2 \, \Omega_e \dot M_{\rm wind} \left[ \frac{B_p^2R_*^2}{\dot M_{\rm wind}\sqrt{K_2^2 v_{\rm esc}^2 + \Omega_e^2 R_*^2}} \right]^{2\,m} R_*^2.
\end{equation}
In the equation above, $v_{\rm esc}^2 = 2GM_*/R_*$ is the escape velocity at the stellar surface, $B_p$ is the surface magnetic field and $\dot M_{\rm wind}$ is the stellar mass loss rate due to the wind.
In the equation above, the constants $K_1$, $K_2$, and $m$ are fixed according to the numerical simulations of \citet{Matt_etal:2012a}; their values are: $K_1=1.30$, $K_2=0.0506$, $m=0.2177$.
The values of $B_p$ and $\dot M_{\rm wind}$ are calculated using the \texttt{BOREAS} subroutine \citep{Cranmer_Saar:2011} with the filling factor slightly modified in order to reproduce the average filling factor of the present Sun as in \cite{Gallet_Bouvier:2013}.
\item
The braking law proposed by \cite{Matt_etal:2015}.
They derived a physically motivated scaling for the dependence of the wind braking law on Rossby number and adopted an empirical scaling with stellar mass and radius.
For our purposes, the \cite{Matt_etal:2015} braking law can be recast in the form:
\begin{equation}
\label{mbwb}
\dot J_{\rm wb} =
\left\{
\begin{array}{cc}
- K_w {\cal{F}} \left(\dfrac{\tau_*}{\tau_{\odot}}\right)^{p}  \Omega_e^{p+1} & \mathrm{for} \; Ro > Ro_{\odot} / \chi \\
\\
- K_w {\cal{F}} \chi^p \Omega_e & \mathrm{for} \; Ro \le Ro_{\odot} / \chi \\
\end{array}
\right. ,
\end{equation}
with
\begin{equation}
{\cal{F}} \equiv \left(\frac{R_*}{R_{\odot}}\right)^{3.1} 
\left(\frac{M_*}{M_{\odot}}\right)^{0.5} ,
\end{equation}
and
\begin{equation}
\chi \equiv \frac{Ro_{\odot}}{Ro_{\rm sat}}
\equiv \frac{\Omega_{\rm sat} \tau}{\Omega_{\odot} \tau_{\odot}} .
\end{equation}
Following \cite{Matt_etal:2015}, we adopt $p=2$, which implies that this model also reproduces the empirical \citet{Skumanich:1972} law in the solid-body rotation limit.
The saturation regime is equivalent to that introduced by \cite{Krishnamurthi_etal:1997}; here we adopt $\chi=10$ as in \cite{Matt_etal:2015} (see also Sect.\,\ref{sec:results}).
\end{enumerate}

\subsection{MCMC fitting of the TZM parameters}
\label{sec:mcmc_method}

According to the TZM, the rotational evolution is completely determined by the stellar mass, which also selects the appropriate background stellar model, and by five parameters: the initial period $P_0$, the disc lifetime $\tau_{\rm disc}$, the coupling timescale $\tau_{\rm cp}$, and the two parameters contained in the wind braking laws \eqref{kawalerwb} or \eqref{modkawwb}.
We use a MCMC approach to constrain the values of these parameters.

The MCMC method is a powerful technique to obtain quantitative inferences on a set of model parameters $\bf p$ from the observational data $\bf d$ (see, e.g. Appendix~A of \citealt{Tegmark_etal:2004} for a concise introduction).
In the case at hand, the vector of model parameters for the TZM with, e.g, the braking law~\eqref{kawalerwb} is ${\bf p} \equiv (P_0,\tau_{\rm disc}, \tau_{\rm cp},K_w,\Omega_{\rm sat})$, while for each mass listed in Table~\ref{tab:periods} the vector of data points contains the periods from the non-parametric fits at the various cluster ages.
For the models with $M_* = 1.00 \, M_\odot$, we also consider the additional constraint based on the  current solar rotation rate\footnote{We adopt the values $t_\odot = 4.57$ Gyr, $\Omega_\odot = 2.903\cdot 10^{-6}$ s$^{-1}$.}: $\Omega_e(t_\odot)=1.0\pm0.1\, \Omega_{\odot}$.

Formally, Bayes' Theorem relates the probability of the model, given the data, i.e., the {\it posterior probability}\footnote{The standard notation ${\cal P}(A|B)$ denotes the conditional probability of event $A$, given that event $B$ is observed.} ${\cal P}({\bf p}|{\bf d})$, to the compatibility of the data with the model, i.e., the {\it likelihood} ${\cal P}({\bf d}|{\bf p})$, and the intrinsic likeliness of the model (prior probability, or simply {\it prior}), ${\cal P}({\bf p})$:
\begin{equation}
\label{bayes}
{\cal P}({\bf p}|{\bf d}) = \frac{ {\cal P}({\bf d}|{\bf p}) \,{\cal P}({\bf p}) }{{\cal P}({\bf d})}  \propto {\cal L}({\bf d};{\bf p}) \, {\cal P}({\bf p}).
\end{equation}
In the following, we ignore the normalization ${\cal P}({\bf d})$, which is a constant with respect to the models, and we use the notation ${\cal L}({\bf d};{\bf p})$ for the likelihood.
In essence, the MCMC method consists of sampling the posterior probability by means of a quasi-random walk in the space of the parameters $\bf p$, constructed as a chain of jumps guided by the values of the likelihood ${\cal L}({\bf d};{\bf p})$. 
Each step in the chain depends only on the previous one (Markov Chain), and is partly stochastic (Monte Carlo). 
The steps are performed according to the following two-stage rule:
\begin{enumerate}
\item {\it Propose stage}: ${\bf p}_{\rm old} \rightarrow {\bf p}_{\rm trial} = {\bf p}_{\rm old} + \Delta {\bf p}$, with $\Delta {\bf p}$ extracted from the jump function $f(\Delta {\bf p})$;
\item {\it Acceptance/rejection stage}: the probability to accept the jump is given by the Metropolis-Hastings rule \citep{Metropolis:1953,Hastings:1970}:
\begin{equation}
\label{methast}
{\cal P}_{\rm accept} = \min \left[ \dfrac{{\cal L}({\bf d};{\bf p}_{\rm trial})}{{\cal L}({\bf d};{\bf p}_{\rm old})}, 1 \right] .
\end{equation}
\end{enumerate}
The Metropolis-Hastings rule ensures that a trial step increasing the likelihood is always accepted (${\cal L}({\bf d};{\bf p}_{\rm trial})>{\cal L}({\bf d};{\bf p}_{\rm old}) \Rightarrow {\cal P}_{\rm accept}=1$), while at the same time even steps which reduce the likelihood retain a (proportionally small) non-zero probability to be accepted.
This procedure is completely specified by the choice of the functions $f(\Delta {\bf p})$ and ${\cal L}({\bf d};{\bf p})$.

We use a multi-dimensional Gaussian form for the jump function:
\begin{equation}
\label{jumpfun}
f(\Delta {\bf p}) \propto \prod_j e^{\Delta p_j^2/2 \ell_j^2},
\end{equation}
where $\ell_j$ are the jump sizes and the index $j$ (up to $5$ in our case) runs over the elements of the parameter vector $\bf p$.
The jump function is critical to ensure that the steps in the chain are optimally correlated, providing a useful sampling of the posterior probability: the optimal regime is in between $100\%$ step acceptance (i.e., a purely random walk) and $100\%$ step rejection.

We define the likelihood in terms of the chi-square:
\begin{equation}
\label{likelihood}
{\cal L}({\bf d}; {\bf m}) = e^{-\chi^2/2}.
\end{equation}
The chi-square contains contributions from each cluster for which an estimate of the rotation period on the slow-rotator sequence at the mass considered is available:   
\begin{equation}
\label{chisq}
\chi^2 = \sum_{i=1}^n \left[  \frac{ \Omega_{m,i} - \Omega_{o,i} }{\sigma_{\Omega,i}  } \right]^2,
\end{equation}
where $\Omega_{m,i}\equiv\Omega_e(t_i)$ is the angular velocity of the convective envelope calculated from a run of the TZM with the set of parameters $\bf p$, while $\Omega_{o,i}$, $\sigma_{\Omega,i}$ are derived from Table~\ref{tab:periods} as follows:
\begin{eqnarray*}
\Omega_{o,i} = \frac{2\pi}{ P_i};
\ \ \ \ 
\sigma_{\rm \Omega,i} = \Omega_{o,i}  \frac{\sigma_{P,i}}{ P_i}.
\end{eqnarray*}

\begin{figure*}[ht]
\begin{center}
\includegraphics[width=0.48\textwidth]{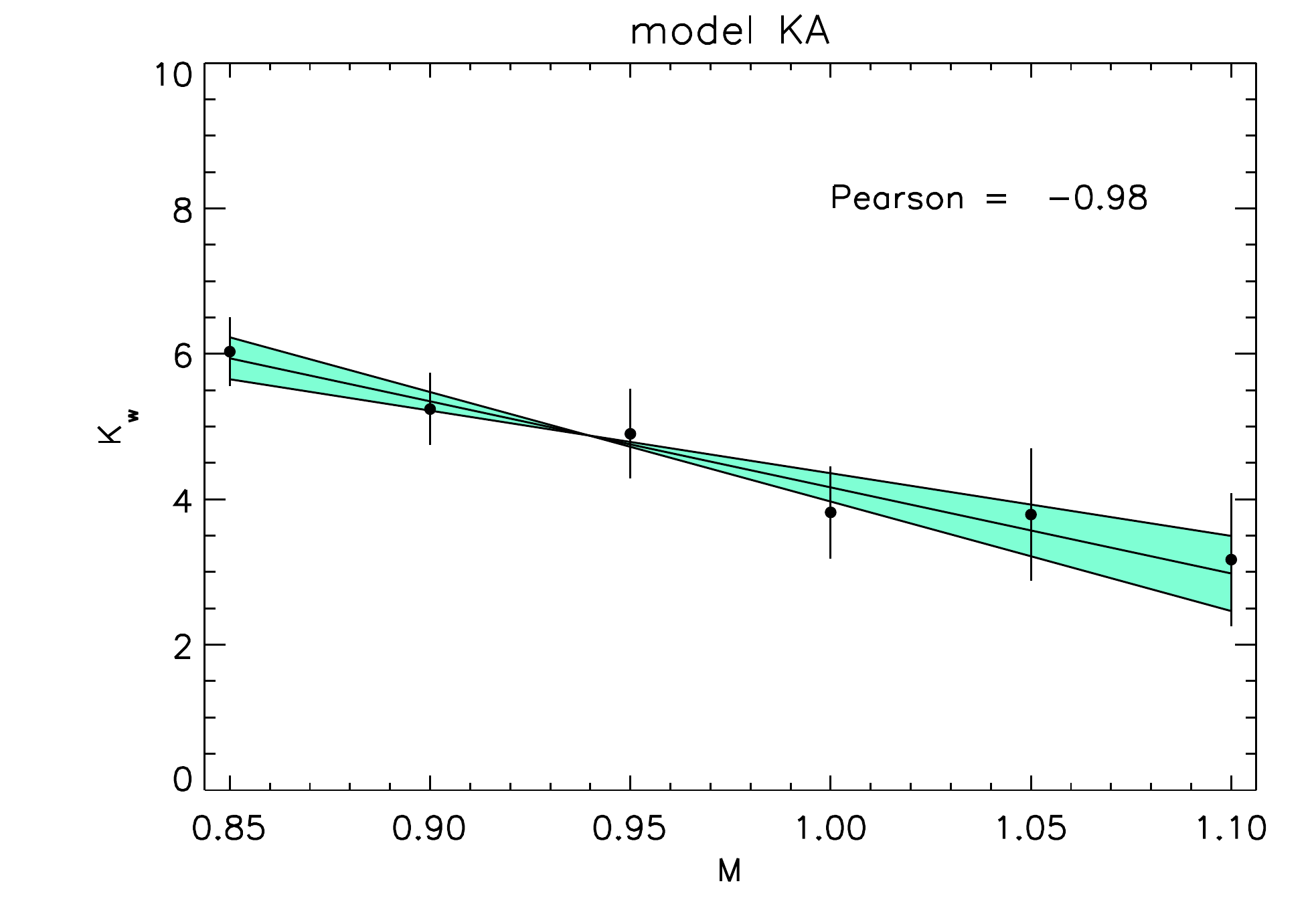}
\includegraphics[width=0.48\textwidth]{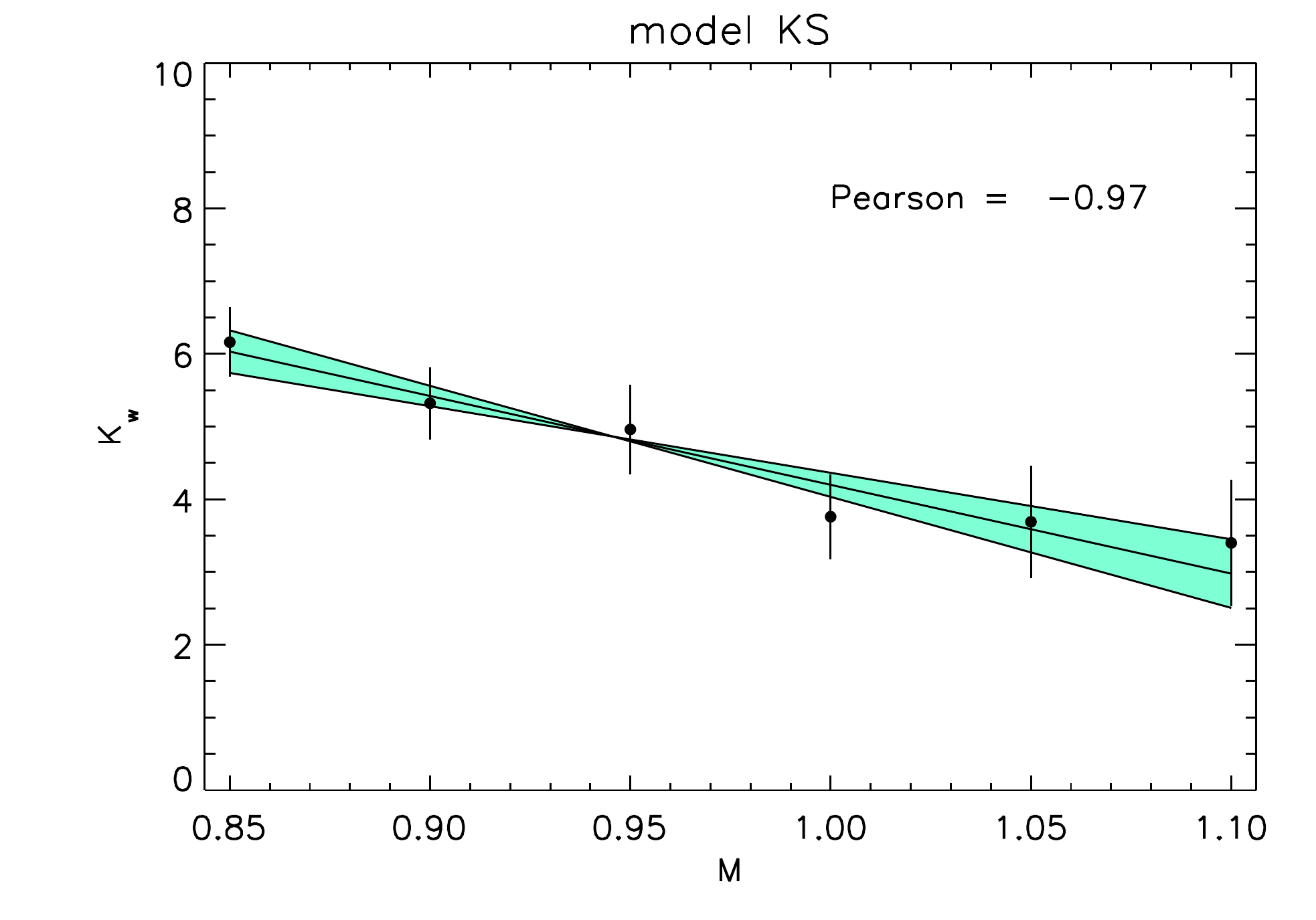}
\includegraphics[width=0.48\textwidth]{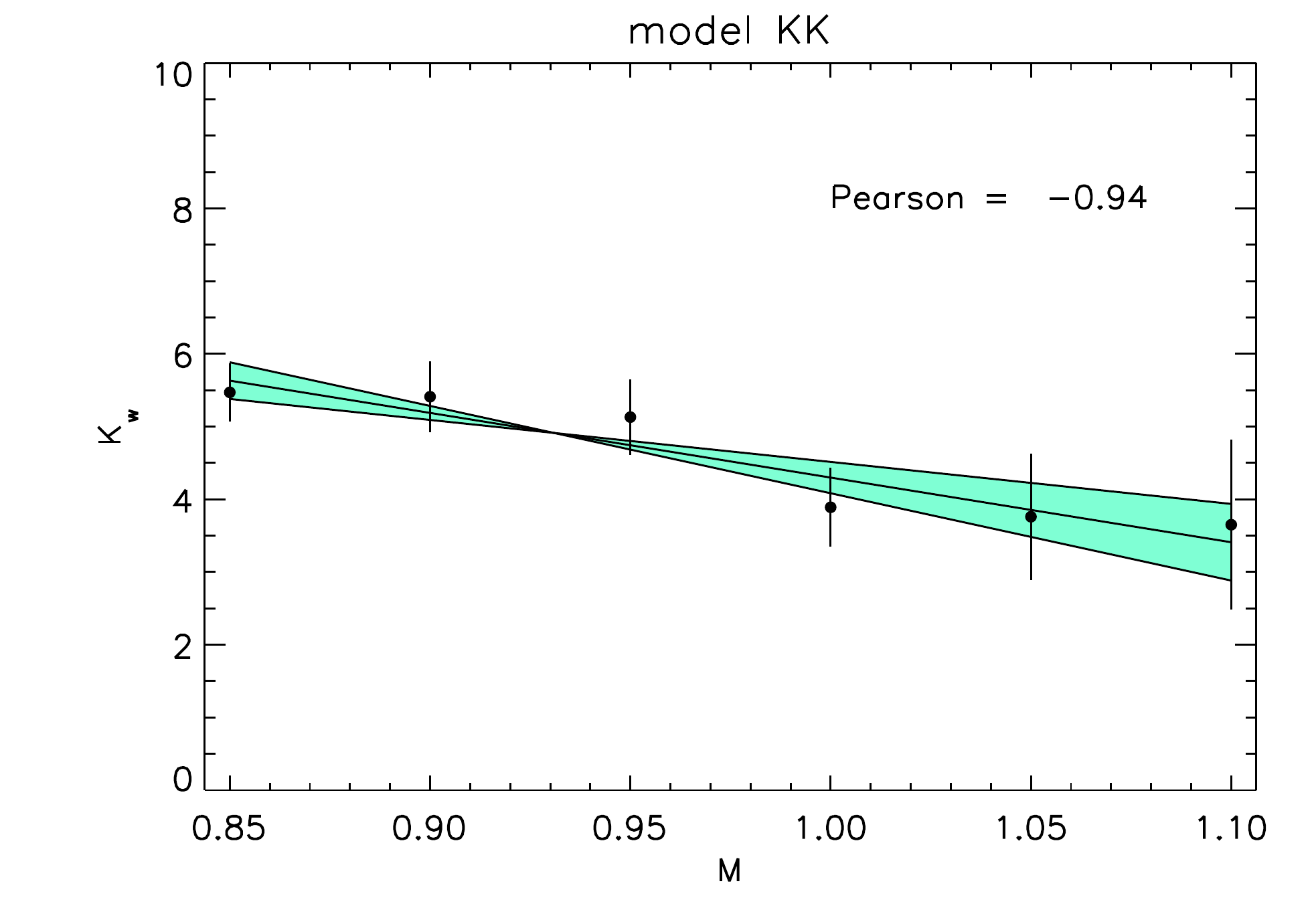}
\includegraphics[width=0.48\textwidth]{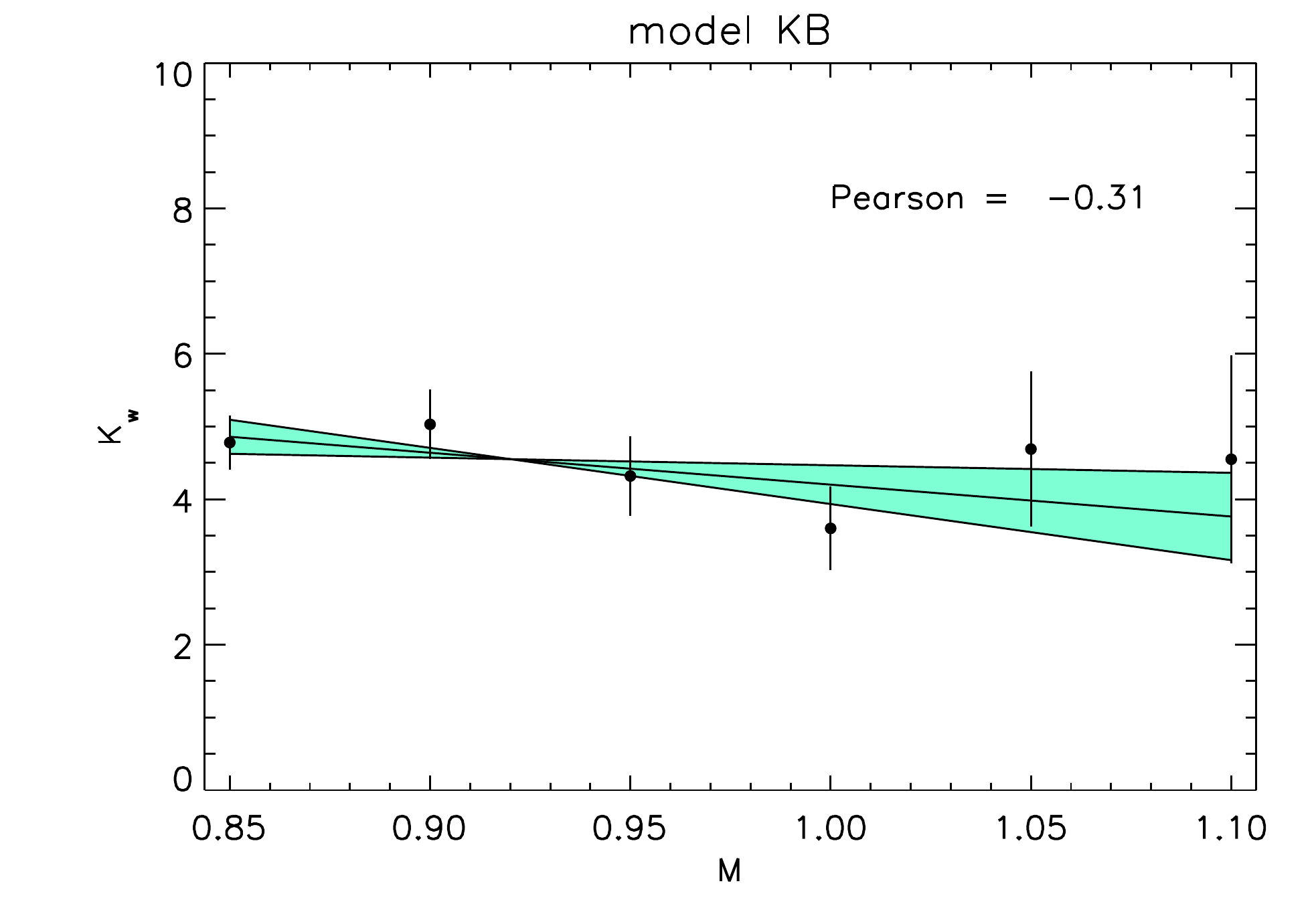}
\includegraphics[width=0.48\textwidth]{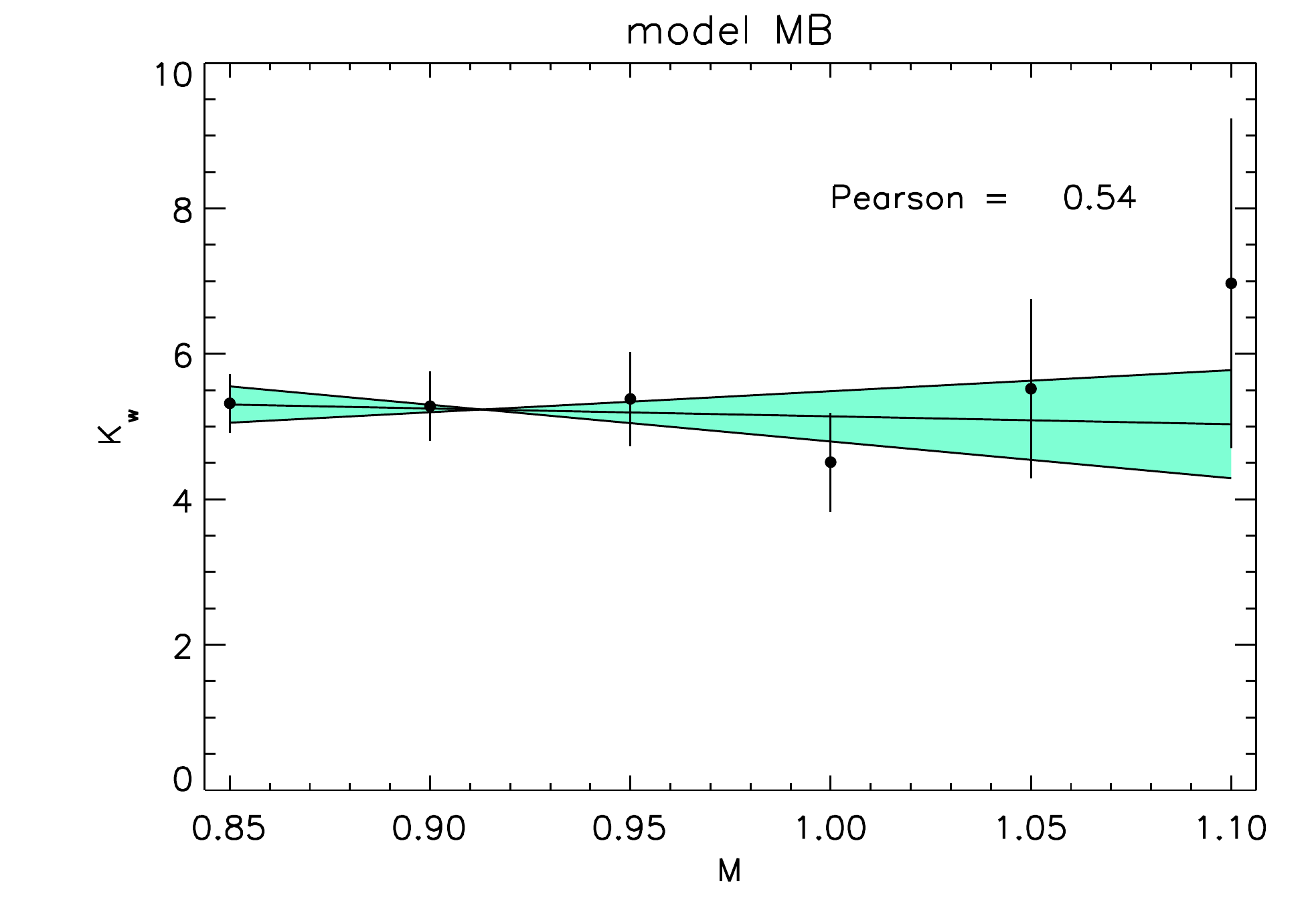}
\includegraphics[width=0.48\textwidth]{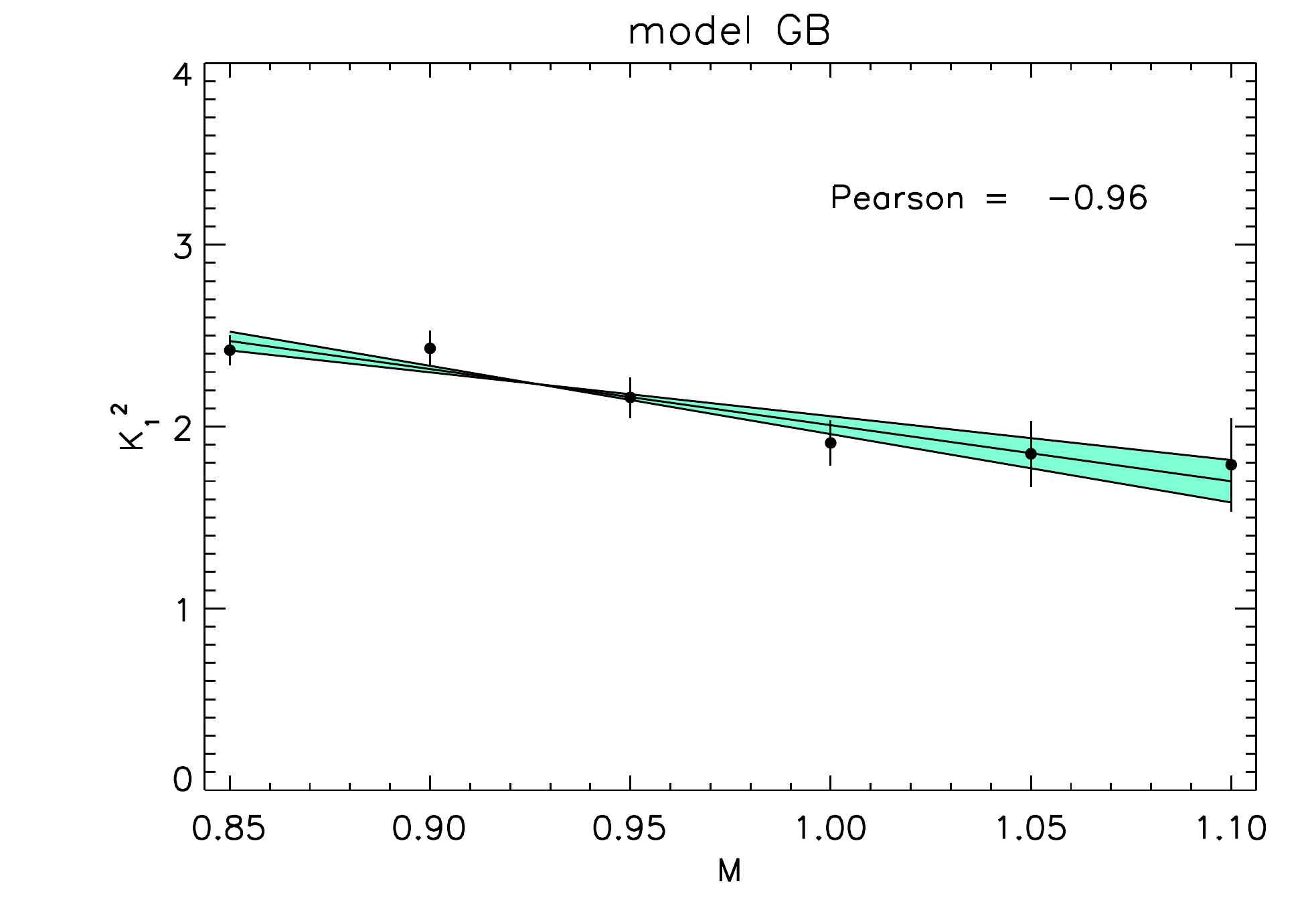}
\caption{Correlation of the parameter $K_w$ for models KA, KS, KK, KB, MB, and $K_1$ for model GB with mass.
The scaling factor of $K_w$ is $1.11 \cdot 10^{47}$\,g\,cm$^2$\,s.
The best linear fit is reported together with the uncertainties on the slope parameter.
The linear Pearson correlation coefficient is given is each panel.}
\label{fig:Kw_vs_mass}
\end{center}
\end{figure*}

In practical applications, running the MCMC procedure at the optimal acceptance regime ($\sim 20\%$ of attempted steps accepted), is often hindered by the existence of correlations in the way the parameters influence the model. 
In our case, for example, we can anticipate the existence of degenerate pairs of values of $P_0$ and $\tau_{\rm disc}$, i.e., a longer initial period and a shorter disc interaction, and vice versa, compensating each other to give a very similar evolution at sufficiently late times (say, a few hundred Myr onwards). 
Apart from computational efficiency, strong correlations can lead to oversampling of the degenerate regions of the parameter space, resulting in very biased, and practically useless, chains.
We deal with this issue as recommended by \citet{Tegmark_etal:2004}, by introducing a transformation from $\bf p$ to an uncorrelated vector $\bf u$, and quantitatively checking the quality of the chain:
\begin{itemize}
\item To transform to uncorrelated variables, we calculate the correlation matrix of the parameters 
$$\tens{C} = \left<{\bf p p^t}\right> - \left<{\bf p}\right>\langle{\bf p^t}\rangle$$
over a portion of the chain (typically, every $100$ steps), then diagonalize it to obtain the matrices $\tens{\Lambda}$ and $\tens{R}$:
\begin{equation*}
\tens{C} = \tens{R}\tens{\Lambda}\tens{R^t};
\end{equation*}
since $\tens{C}$ is by definition real and symmetric, the eigenvalues are real and the matrix $\tens{R}$ is orthogonal.
The linear transformation
\begin{equation*}
{\bf u} = \tens{\Lambda^{-1/2}} \tens{R^t} \left[ {\bf p} - \left<{\bf p}\right> \right]
\end{equation*}
results in the very useful properties: $\left< {\bf u}\right>=0$, $\langle{\bf u} {\bf u^t}\rangle= \tens{1}$, i.e., the variables $u_i$ are (linearly) uncorrelated and normalized to one.
Operationally, the vector $\bf p$ is transformed into $\bf u$ before substitution into the jump function \eqref{jumpfun}; the trial step is performed in the variables $\bf u$, and then transformed back if the step is accepted.
Since $\bf u$ is normalized to one, the choice of the jump sizes is particularly simple: $\ell_j=\ell \approx 1$.
The correlation matrix $\tens{C}$ is recalculated, and the matrices $\tens{\Lambda}$ and $\tens{R}$ are updated, every $\sim 100$ steps.
\item To evaluate the quality of the chain, we calculate the autocorrelation of each parameter $p_j$,
\begin{equation*}
c_j(k) = \frac{\langle p_j(i+k)p_j(i)\rangle - \langle p_j(i) \rangle^2 }{\langle p_j^2(i) \rangle - \langle p_j(i) \rangle^2 }, 
\end{equation*}
where $p_j(i)$, etc., is the value of $p_j$ at the step $i$ of the chain. 
Typically, $c_j(k)$ decreases with $k$ (its value at $k=1$ being unity by definition): we define the correlation length of the chain as the value $\bar k_j$ for which $c_j(k)$ drops below $0.5$ for the first time. 
The {\it effective length} of the chain $\bar N_j$ is thus the number of steps $N$ divided by $\bar k_j$:
\begin{equation*}
\bar N_j = {N}/{\bar k_j}.
\end{equation*}
This is a measure of the number of independent points in the chain: if $\bar N_j$ is sufficiently large, the average of $p_j$ over the chain, along with its standard deviation, can be considered representative of the best-fitting value of $p_j$ and its uncertainty, respectively.
To ensure that we only use a portion of the chain sufficiently close to convergence, the first few hundred steps ({\it burn-in phase}) are discarded.
\end{itemize}

\begin{figure}
\begin{center}
\includegraphics[width=0.48\textwidth]{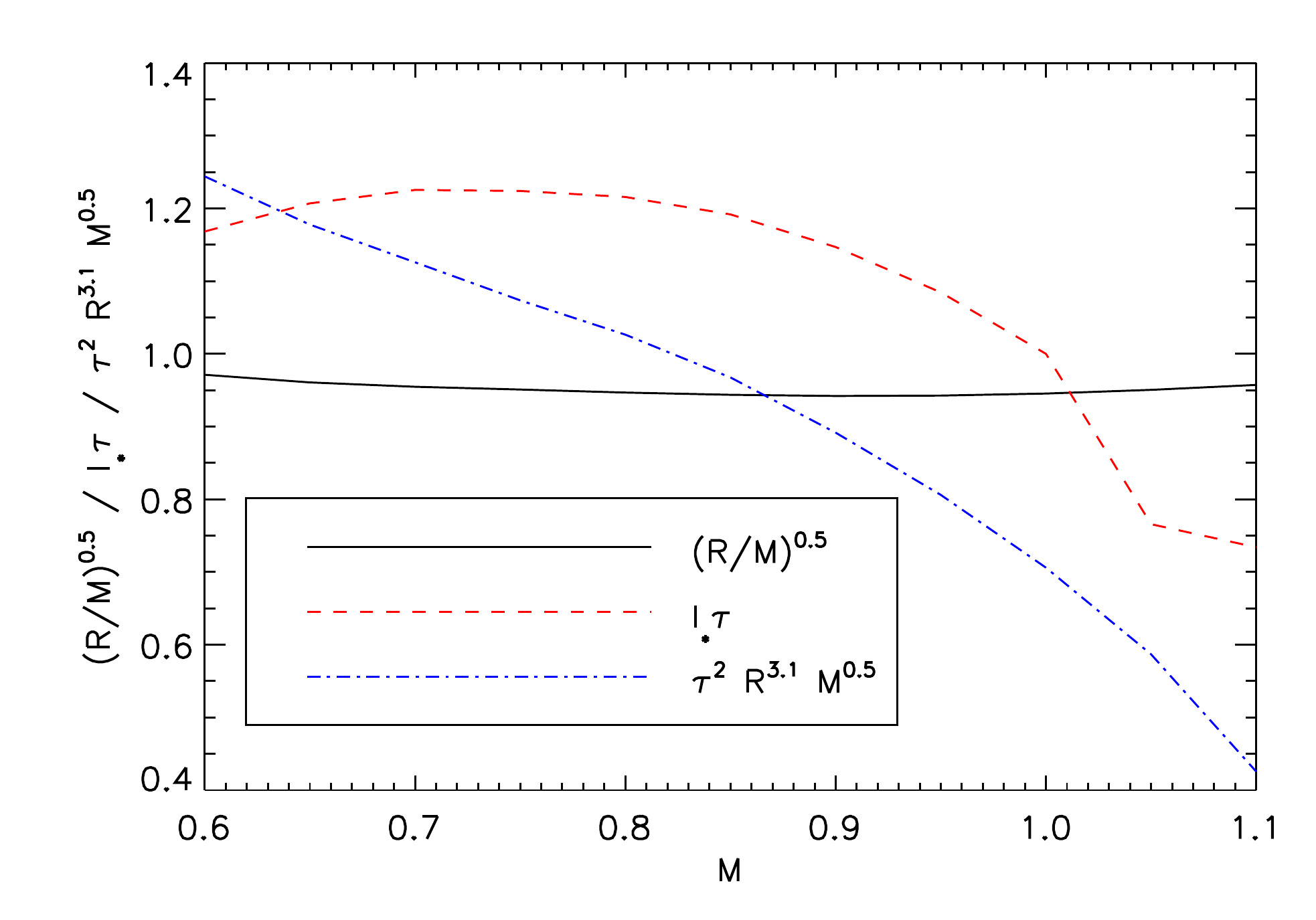}
\caption{
Comparison of the mass dependence of models KA (solid black line), KB (red dashed line), and MB (blue dot-dashed line) at $t=0.55$ Gyr. All quantities are in solar units.
}
\label{fig:mass_dependence}
\end{center}
\end{figure}

\begin{table*}[ht]
\caption{Slow-rotator sequence two-zone Monte Carlo Markov-chain fitting results (see text for details on the different models). 
For each mass and model this table reports the resulting $\chi^2$, the fitted $K_w$ for models KA, KS, KK, KB, and MB or $K_1^2$ for model GB, and the fitted core-envelope coupling timescale $\tau_{\rm cp}$. For both $K$ and $\tau_{\rm cp}$ the table also gives the uncertainty and the effective chain length.
The scaling factor of $K_w$ is $1.11 \cdot 10^{47}$\,g\,cm$^2$\,s.}
\begin{center}
\begin{tabular}{ccccccccc}
\hline
\hline
 Mass ($M_\odot$)  & Model & $\chi^2$ & $K$ & $\sigma_{K}$ & $\bar N_{K}$ & $\tau_{\rm cp}$ (Myr) & $\sigma_{\tau_{\rm cp}}$ (Myr) & $\bar N_{\tau_{\rm cp}}$ \\
 \hline
\\
\multirow{6}{*}{0.85} 
&    KA   &     2.00   &     6.03   &     0.48   &    12250   &    85.5    &    11.1    &     8166 \\
&    KS   &     2.03   &     6.16   &     0.48   &    12250   &    86.8    &    10.7    &     8166 \\
&    KK   &     1.72   &     5.47   &     0.40   &    12250   &    79.7    &    10.6    &     8166 \\
&    KB   &     1.99   &     4.78   &     0.37   &    12250   &    84.6    &    10.8    &     8166 \\
&    MB   &     2.12   &     5.32   &     0.41   &     8166   &    83.1    &    11.5    &     8166 \\
&    GB   &     1.54   &     2.42   &     0.08   &    12250   &    85.0    &     9.1    &     8166 \\
\\
\multirow{6}{*}{0.90}  
&    KA   &     1.21   &     5.24   &     0.50   &     8166   &    51.3    &     9.4    &     8166 \\
&    KS   &     1.25   &     5.32   &     0.50   &     8166   &    52.6    &     9.4    &     8166 \\
&    KK   &     1.28   &     5.41   &     0.49   &     8166   &    53.5    &     9.1    &     8166 \\
&    KB   &     1.61   &     5.03   &     0.48   &     8166   &    59.6    &     9.3    &     8166 \\
&    MB   &     1.44   &     5.28   &     0.48   &     8166   &    52.3    &     9.9    &     8166 \\
&    GB   &     1.09   &     2.43   &     0.10   &    12250   &    67.4    &     7.1    &    12250 \\
\\
 \multirow{6}{*}{0.95}  
&    KA   &     1.05   &     4.90   &     0.61   &     8166   &    34.9    &     9.4    &     8166 \\
&    KS   &     1.11   &     4.96   &     0.62   &     8166   &    35.5    &     9.1    &     8166 \\
&    KK   &     1.45   &     5.13   &     0.52   &     8166   &    38.3    &     7.9    &     8166 \\
&    KB   &     1.07   &     4.32   &     0.54   &     8166   &    35.1    &     9.1    &     8166 \\
&    MB   &     1.11   &     5.38   &     0.65   &     8166   &    34.9    &     9.4    &     8166 \\
&    GB   &     1.14   &     2.16   &     0.11   &     8166   &    49.0    &     6.4    &     8166 \\
\\
\multirow{6}{*}{1.00} 
&    KA   &     1.81   &     3.82   &     0.64   &     8166   &    19.1    &     8.5    &     8166 \\
&    KS   &     1.75   &     3.76   &     0.58   &     8166   &    19.0    &     8.1    &     8166 \\
&    KK   &     1.75   &     3.89   &     0.54   &     8166   &    20.0    &     7.4    &     8166 \\
&    KB   &     1.62   &     3.60   &     0.58   &     8166   &    18.7    &     8.1    &     8166 \\
&    MB   &     1.30   &     4.51   &     0.68   &     8166   &    17.3    &     8.3    &     8166 \\
&    GB   &     2.11   &     1.91   &     0.13   &     8166   &    34.1    &     5.7    &     8166 \\
\\
\multirow{6}{*}{1.05} 
&    KA   &     1.87   &     3.79   &     0.91   &     8166   &    15.6    &     7.9    &     8166 \\
&    KS   &     1.90   &     3.69   &     0.77   &     8166   &    15.6    &     7.5    &     8166 \\
&    KK   &     1.81   &     3.76   &     0.87   &     8166   &    15.4    &     7.8    &     8166 \\
&    KB   &     1.73   &     4.69   &     1.07   &     8166   &    15.3    &     7.4    &     8166 \\
&    MB   &     1.71   &     5.52   &     1.23   &     8166   &    15.0    &     8.3    &     8166 \\
&    GB   &     2.37   &     1.85   &     0.18   &     8166   &    23.0    &     5.2    &     8166 \\
\\
\multirow{6}{*}{1.10} 
&    KA   &     3.45   &     3.17   &     0.91   &     6125   &    12.0    &     6.6    &     6125 \\
&    KS   &     3.52   &     3.40   &     0.87   &     6125   &    12.8    &     6.4    &     6125 \\
&    KK   &     3.61   &     3.65   &     1.17   &     6125   &    12.9    &     7.3    &     6125 \\
&    KB   &     3.25   &     4.55   &     1.43   &     6125   &    12.5    &     6.9    &     6125 \\
&    MB   &     3.29   &     6.97   &     2.27   &     6125   &    12.4    &     7.9    &     6125 \\
&    GB   &     2.29   &     1.87   &     0.28   &     8166   &    14.1    &     4.3    &     8166 \\
\hline
\end{tabular}
\end{center}
\label{tab:mcmc_results}
\end{table*}

\section{Results and discussion}
\label{sec:results}

The empirical description of Sect.\,\ref{sec:empirical} implies that, once a star settles on the slow-rotator sequence, it forgets most of its earlier rotational history.
As suggested by \cite{Barnes:2003} or \cite{Brown:2014}, fast rotators may have a different magnetic configuration than slow rotators and migrate to the slow-rotator sequence after a change in their magnetic configuration.
This scenario, which still lacks clear observational support \citep[but see][]{Garraffo_ea:2015}, implies that stars may settle on the slow-rotator sequence either after an initial rotational evolution like those described in Sect.\,\ref{sec:tzm} and \ref{secwb}, with parameters constant in time, or after an entirely different one, in which the parameters change with surface magnetic field configuration \citep[e.g.][]{Brown:2014}.
In the first case, convergence on the slow-rotator sequence would be just a consequence of the $\Omega$ dependence of the wind braking law, including the effects of the saturation regime, which is implicitly assumed in building our starting conditions.
In the second case, both the overall scale factor and the exponent of $\Omega$ in the wind braking law may change at some stage of the evolution, which would require a different approach than that adopted here.
In both cases, however, once on the slow-rotator sequence, the subsequent rotation period evolution is manifestly independent of the previous history, although the abundance of light elements may still depend on it \citep[see, e.g.][]{Bouvier:2008}.

Our fitting, therefore, cannot constrain the stellar rotational history before the settling on the slow-rotator sequence.
In particular, it cannot constrain $P_0$ and $\tau_{\rm disc}$ even if the earlier rotational history follows one of the models listed in Sect.\,\ref{sec:tzm} and \ref{secwb} with parameters constant in time. 
Nevertheless, such parameters are necessary in our modeling for advancing the rotational evolution from the birth-line to the slow-rotator sequence.
We therefore treat $P_0$ and $\tau_{\rm disc}$ as nuisance parameters with reasonable priors and marginalize them out.
The prior on $P_0$ is uniformly distributed over the observed period range in the ONC \citep{Rebull_etal:2004}.
The prior on $\tau_{\rm disc}$ is assumed to be an exponential decay with an $e$-folding time of $\approx 5$ Myr \citep[see, e.g. Fig.\,11 in][]{Hernandez_etal:2008}.

In our modeling, $P_0$ and $\tau_{\rm disc}$ are degenerate with respect to the slow-rotator sequence evolution, in the sense that long $P_0$ and short $\tau_{\rm disc}$ can lead to the same period at the age of the Pleiades as short $P_0$ and long $\tau_{\rm disc}$.
Therefore, after checking compatibility by letting all parameters vary, we set $\tau_{\rm disc} = 3$\,Myr and marginalize $P_0$ out for each mass and model.
It is worth stressing that this is just a convenient way of setting the initial conditions for the slow-rotator sequence evolution and that no inference can be made on the actual distribution of $P_0$ and $\tau_{\rm disc}$.

The duration of the phase of the early stellar evolution that is affected by the saturation regime (Eq.\,\ref{kawalerwb} or \ref{mbwb}) depends on the assumptions made on the saturation threshold.
As a consequence, the star may leave the saturation regime before or significantly after its settling on the slow-rotator sequence.
In fact, if we consider the relationship proposed by \cite{Krishnamurthi_etal:1997}:
\begin{equation}
\label{eq:Krishna}
\Omega_{\rm sat} =  \frac{\tau}{\tau_{\odot}} \Omega_{\rm sat \odot},
\end{equation}
with $\Omega_{\rm sat \odot} = 10\, \Omega_{\odot}$, or, equivalently, the saturation regime in the formulation of \cite{Matt_etal:2015} (see Sect.\,\ref{secwb}), stars of approximately solar mass will leave the saturation regime before settling on the slow-rotator sequence (starting from the Pleiades age in our modeling), but stars of lower mass will stay in the saturation regime for a significant part of their earlier evolution on the slow-rotator sequence.
For this reasons, we adopt the strategy of testing different assumptions on $\Omega_{\rm sat}$, rather than trying to fit it as a free parameter.
We therefore consider Eq.\,\eqref{kawalerwb} with: 
\begin{itemize}
\item no saturation (model KA);
\item $\Omega_{\rm sat} = 10\, \Omega_{\odot}$ for all masses (model KS);
\item $\Omega_{\rm sat}$ as in Eq.\,\eqref{eq:Krishna} (model KK).
\end{itemize}
Note that an implicit mass dependence (through $\tau$) enters both model KK, as a $\tau$-dependent saturation threshold, and Eq.\,\eqref{modkawwb} (hereafter model KB), as a factor in the wind angular momentum loss. 
On the other hand, Eq.\,\eqref{mbwb} (hereafter model MB) contains a complex mass dependence that is both explicit and implicit, through the dependence on $\tau$ and $R_*$, and the $\tau$-dependence of the saturation threshold, respectively.

\begin{figure}[ht]
\begin{center}
\includegraphics[width=0.48\textwidth]{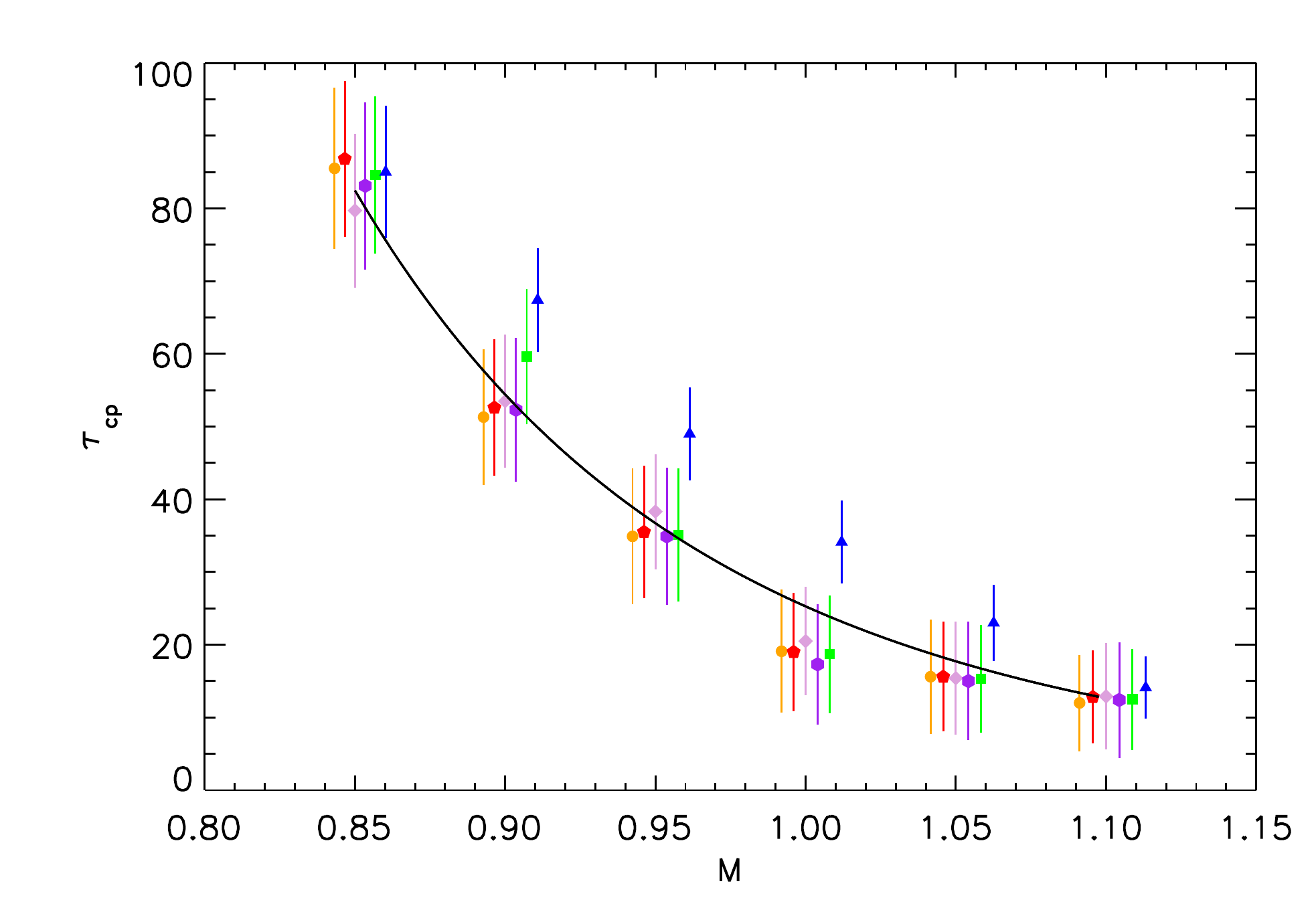}
\caption{
$\tau_{\rm cp}$ vs. $M$ for models KA (orange circle), KS (red pentagon), KK (plum diamond), MB (purple hexagon), KB (green square) and GB (blue triangle).
Results from different models at the same mass are slightly shifted in abscissa for readability.
The solid line represents the relation $\tau_{\rm cp} \propto M^{-7.28}$ obtained from model KB with an average $\langle K_w \rangle = 4.5$. 
}
\label{fig:tauc_vs_mass}
\end{center}
\end{figure}

The wind braking law \eqref{galletwb} (model GB hereafter) contains, in principle, no adjustable parameters.
However, in order to compare it with models derived from Eqs.\,\eqref{kawalerwb}, \eqref{modkawwb}, and \eqref{mbwb}, where $K_w$ is treated as a free parameter, we assume $K_1$ variable and check whether the fitted $K_1$ shows some residual correlation with mass.

By fitting $K_w$ or $K_1$, we both test how well the models describe the mass dependence of the rotational evolution, and correct the mass dependence using the observational data.

\begin{figure*}[ht]
\begin{center}
\includegraphics[width=0.44\textwidth]{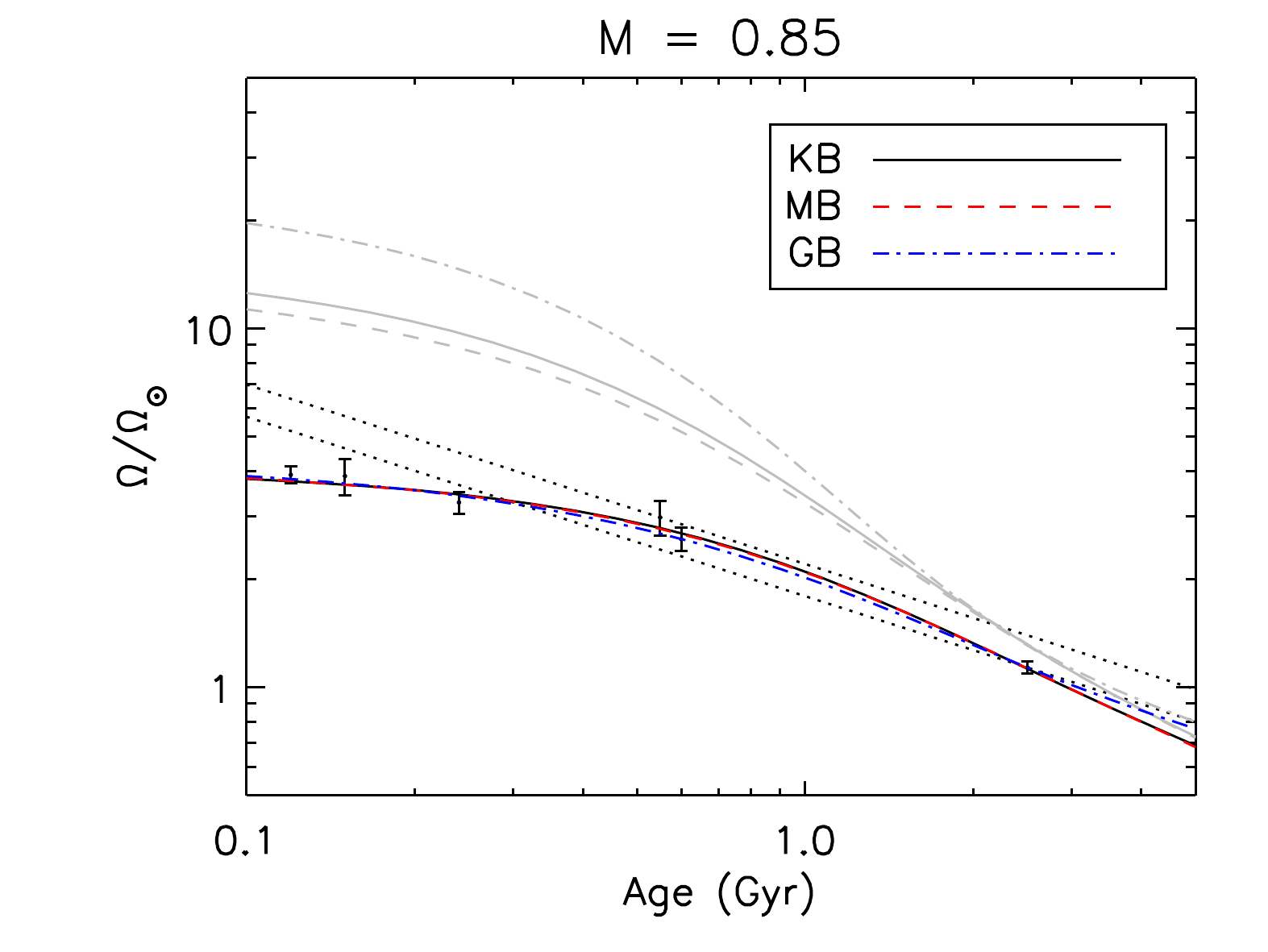}
\includegraphics[width=0.44\textwidth]{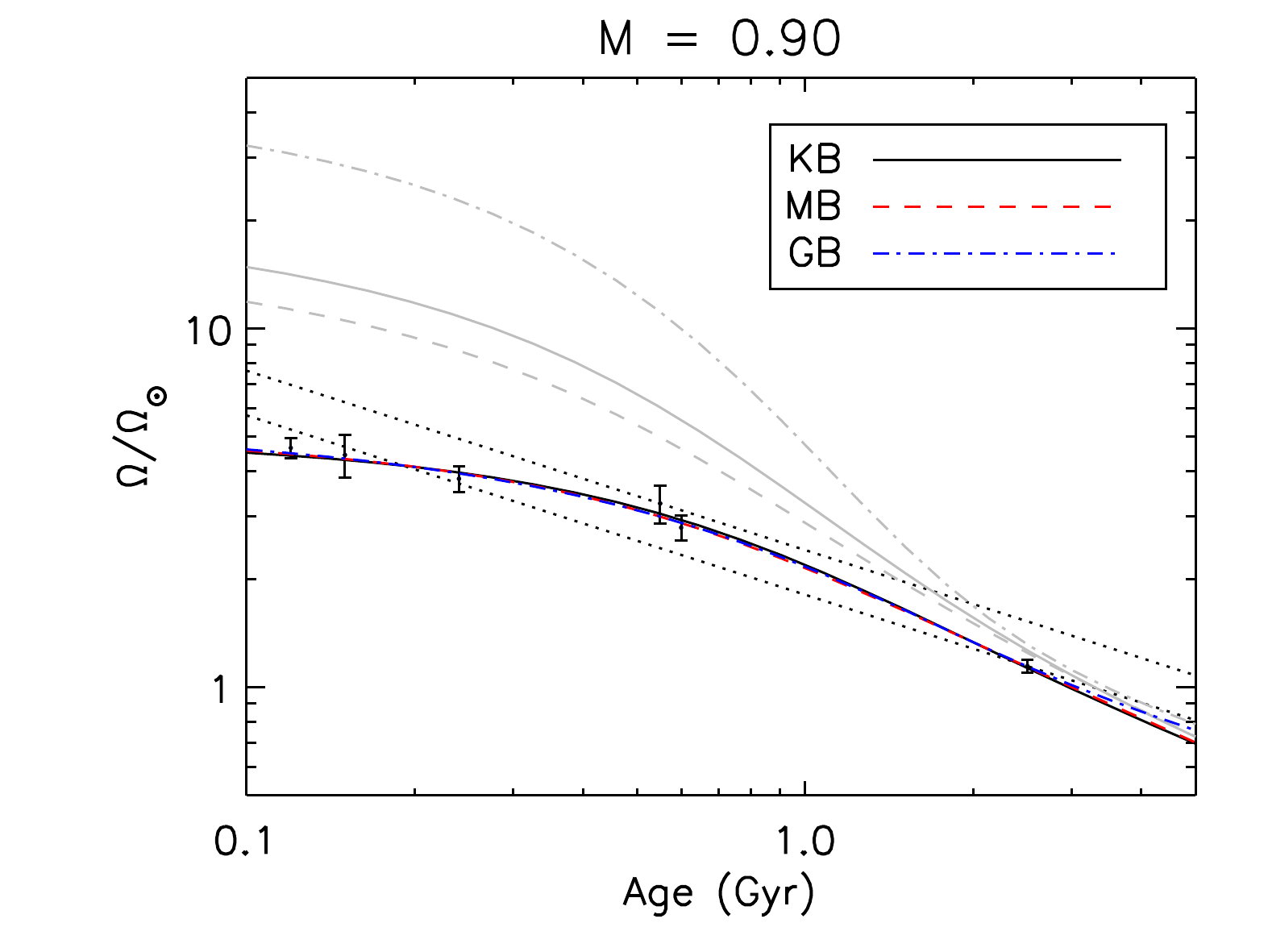}
\includegraphics[width=0.44\textwidth]{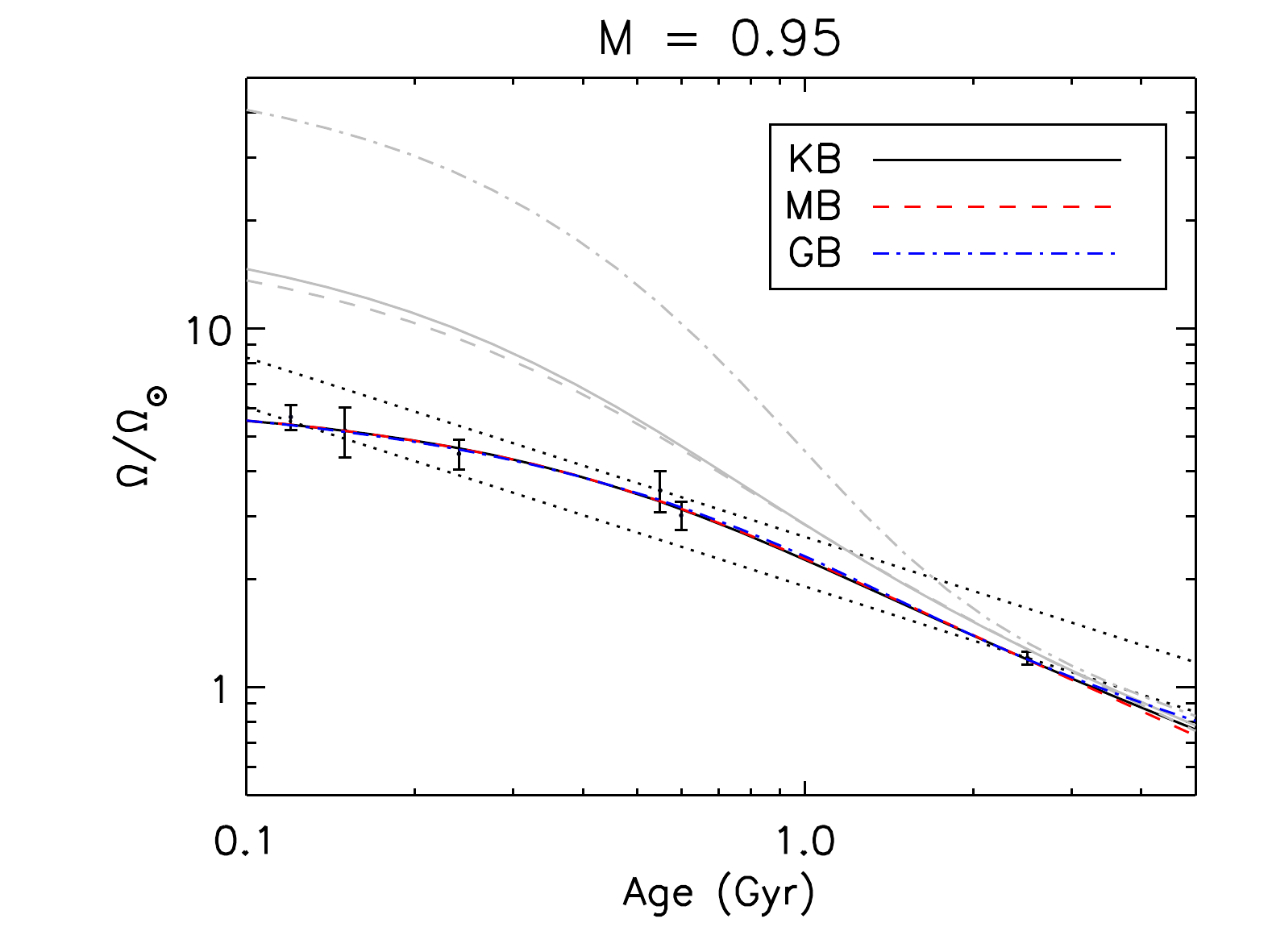}
\includegraphics[width=0.44\textwidth]{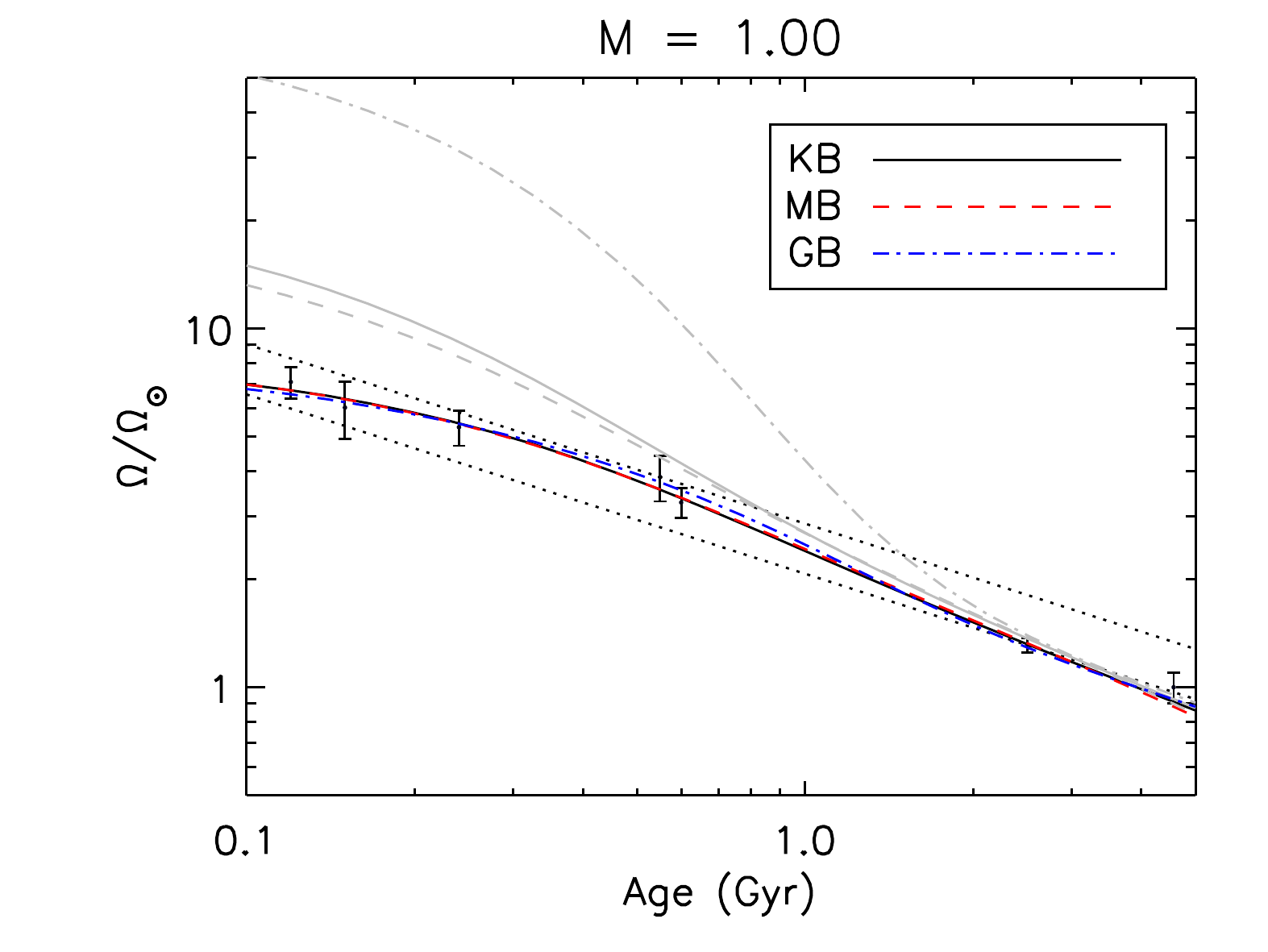}
\includegraphics[width=0.44\textwidth]{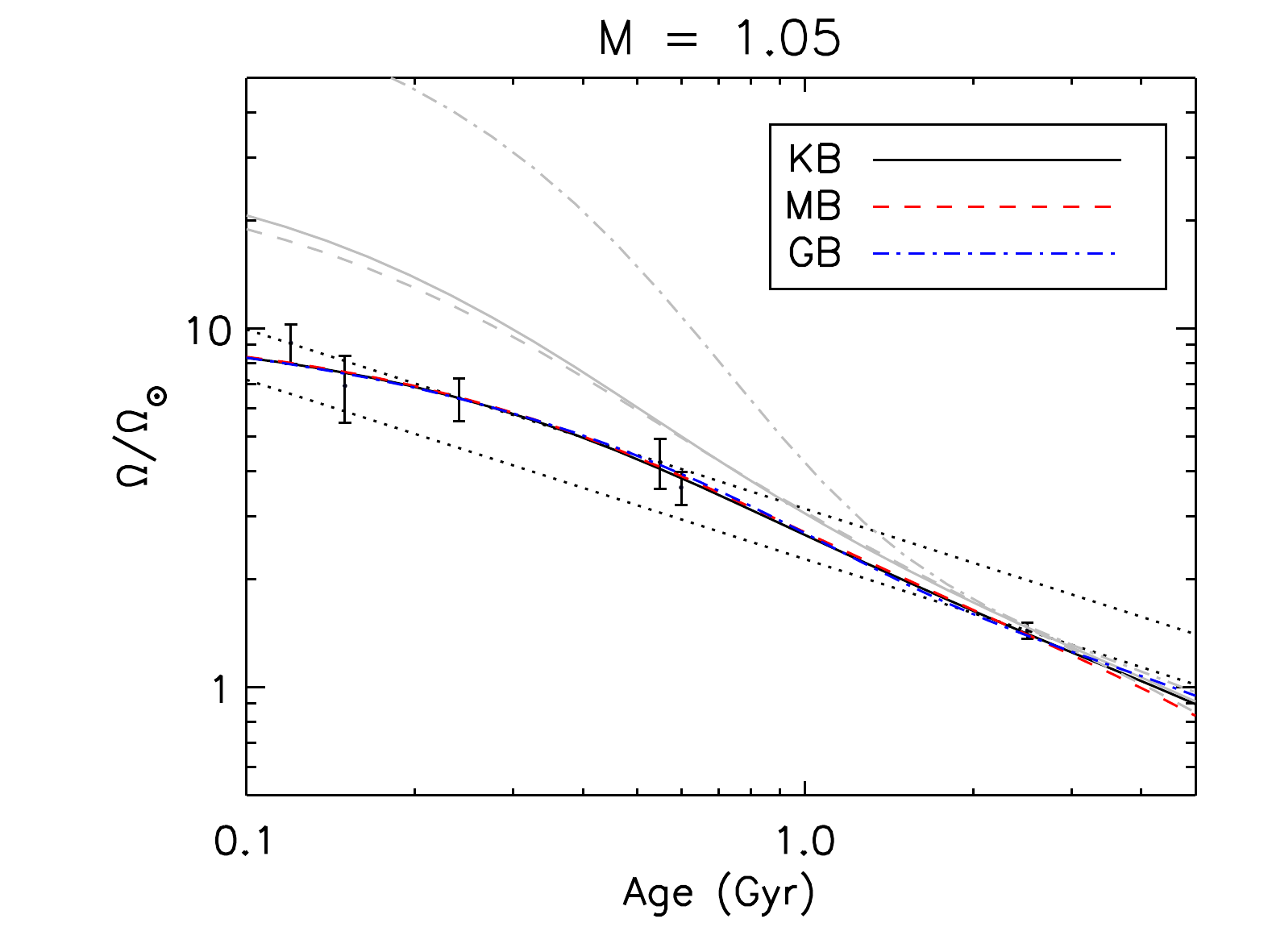}
\includegraphics[width=0.44\textwidth]{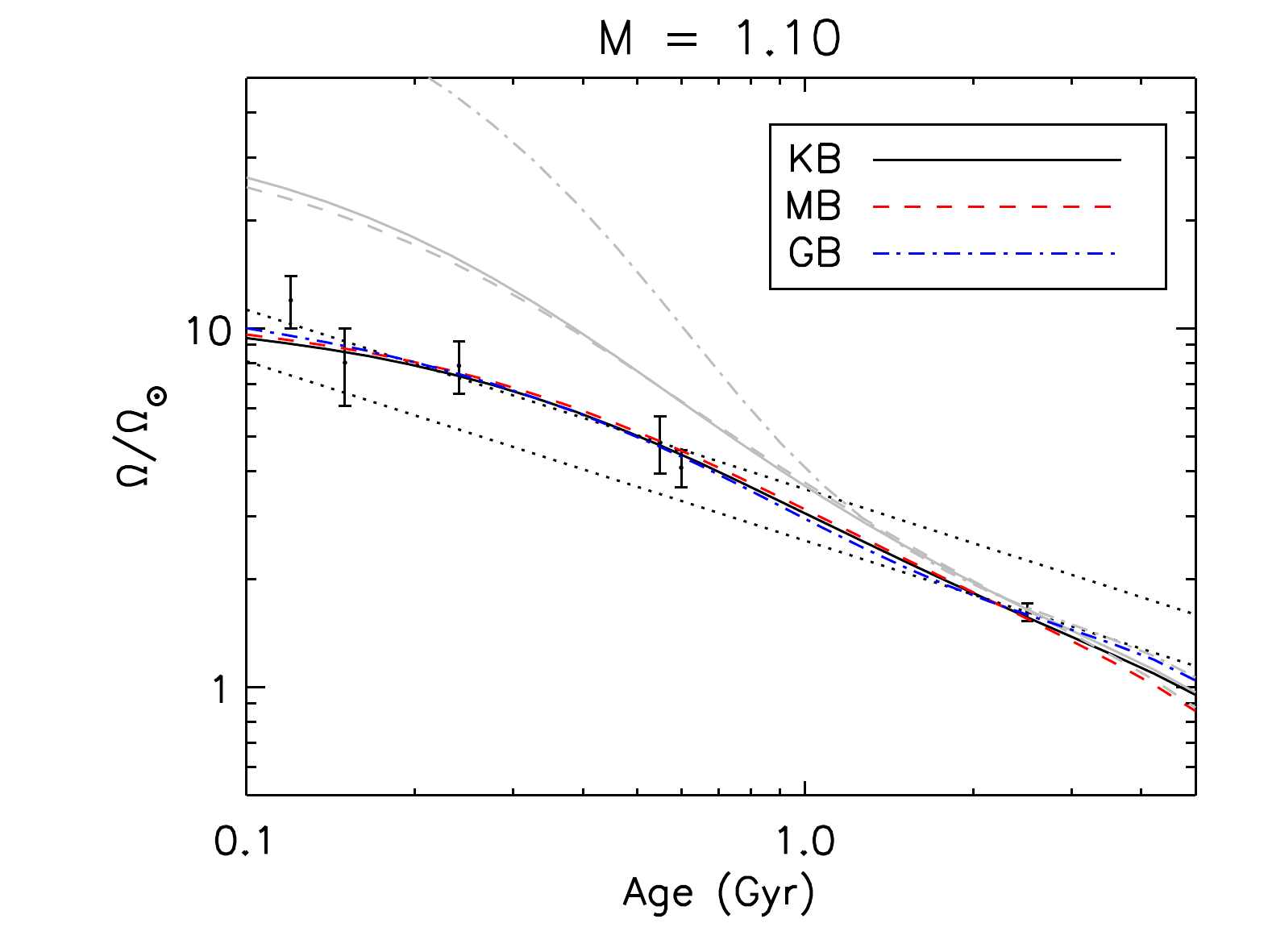}
\caption{
Comparison of the stellar envelope angular velocity evolution according to models KB (black solid line), MB (red dashed line), and GB (blue dot-dashed line) with observations. 
Mass scaling is corrected according to the MCMC fit for all models.
The corresponding core angular velocity for each model is reported in grey with the same line style as for the envelope angular velocity.
}
\label{fig:tracks}
\end{center}
\end{figure*}

After setting the initial conditions, we are left with two free parameters, ${\bf p} = (\tau_{\rm cp},K_w)$ for the models KA, KS, KK, KB, and MB and ${\bf p} = (\tau_{\rm cp},K_1)$ for model GB.
We fit these parameters using the MCMC method described in Sect.\,\ref{sec:mcmc_method} and the data of Table~\ref{tab:periods}.
From this data set we exclude data for stellar mass $<0.85\,M_{\odot}$, since, for our purposes, the sampling in age in this mass range is still insufficient\footnote{Note that the strong $\tau_{\rm cp}$ mass dependence discussed in the following implies that, in order to constrain the TZM parameters at $M<0.85\, M_\odot$, data at ages $\gg 0.6$\,Gyr would be required, which is not available yet.}.
We also exclude the NGC6811 periods, although we verify a posteriori (see below) the consistency of our results with the NGC6811 data.
The exclusion of the NGC6811 periods is due to the irregular behavior in the empirical description of the evolution of the slow-rotator sequence of this cluster in comparison to the general trend. 
Indeed, Fig.\,\ref{fig:GyA_periods} shows that the slow-rotator sequence of NGC6811 tends to bend downwards between $0.8$ and $0.9$ $M_{\odot}$, but this trend is not subsequently seen in the NGC6819 periods (at the age of $\approx 2.5$\,Gyr).
As this may outline some remaining problems with the NGC6811 data, which might result in introducing biases in our results, we choose not to use the NGC811 data in the fitting, but we verify a posteriori the compatibility of the final results with this data.

\begin{figure*}[hp]
\begin{center}
\includegraphics[width=0.44\textwidth]{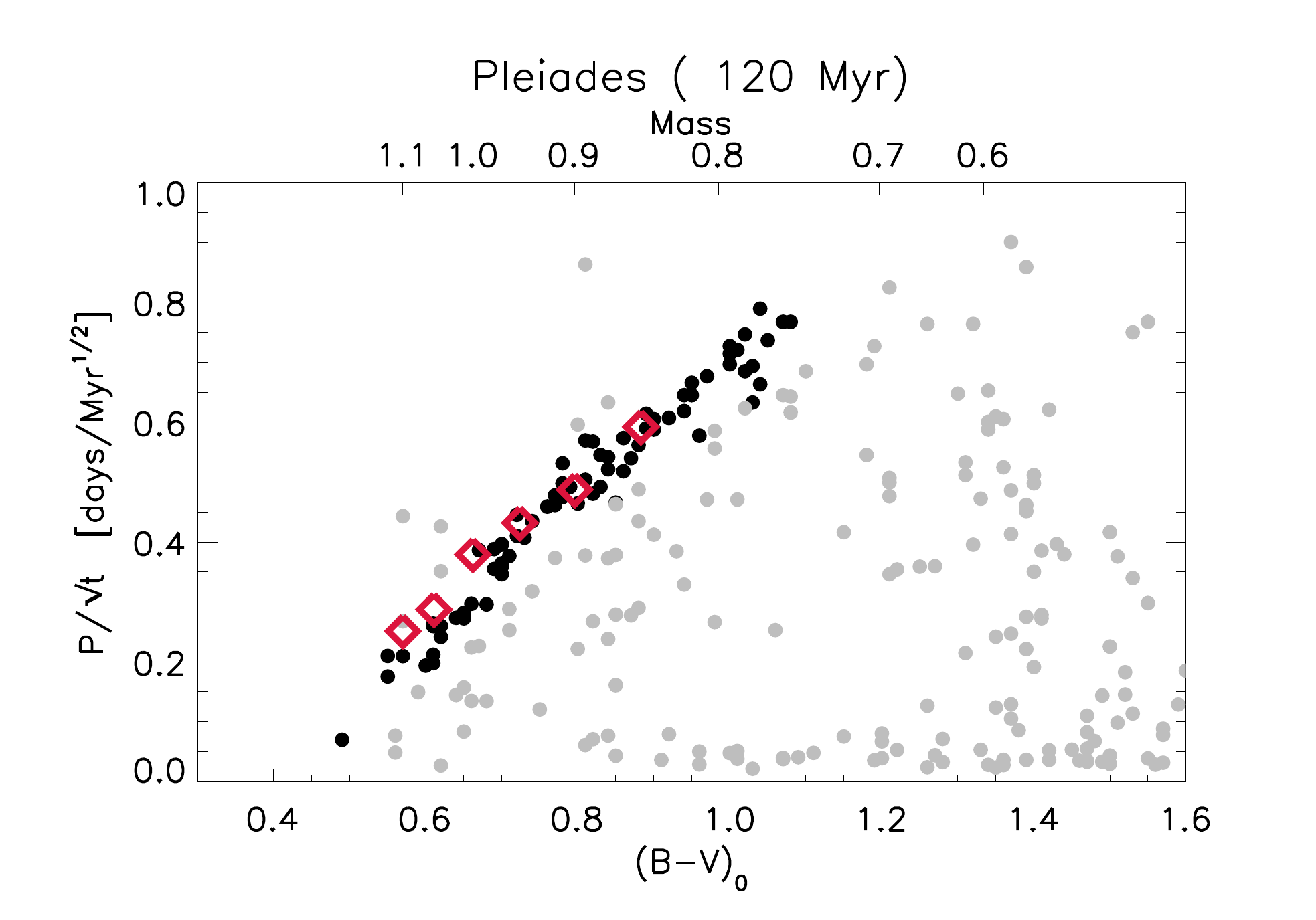}
\includegraphics[width=0.44\textwidth]{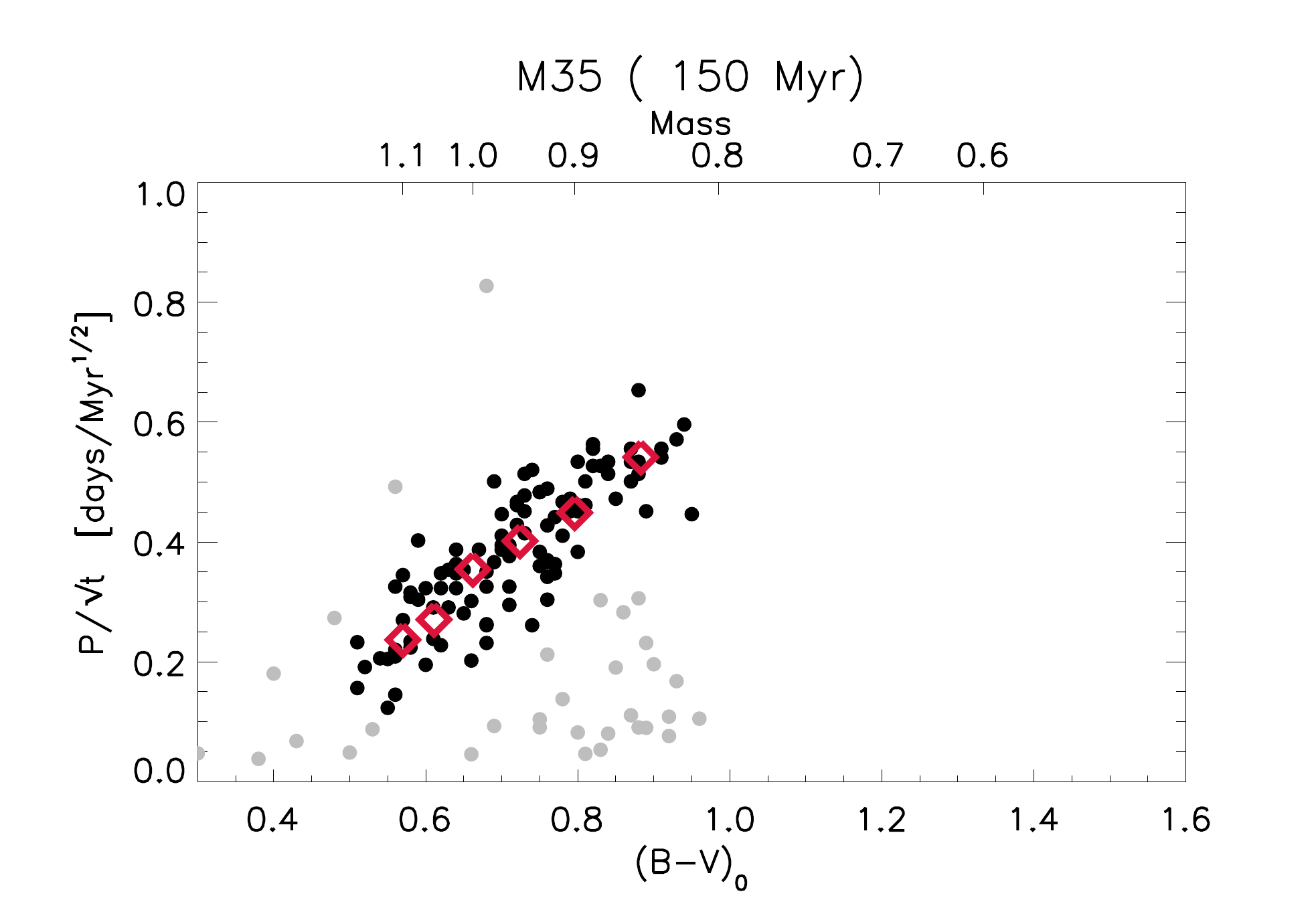}
\includegraphics[width=0.44\textwidth]{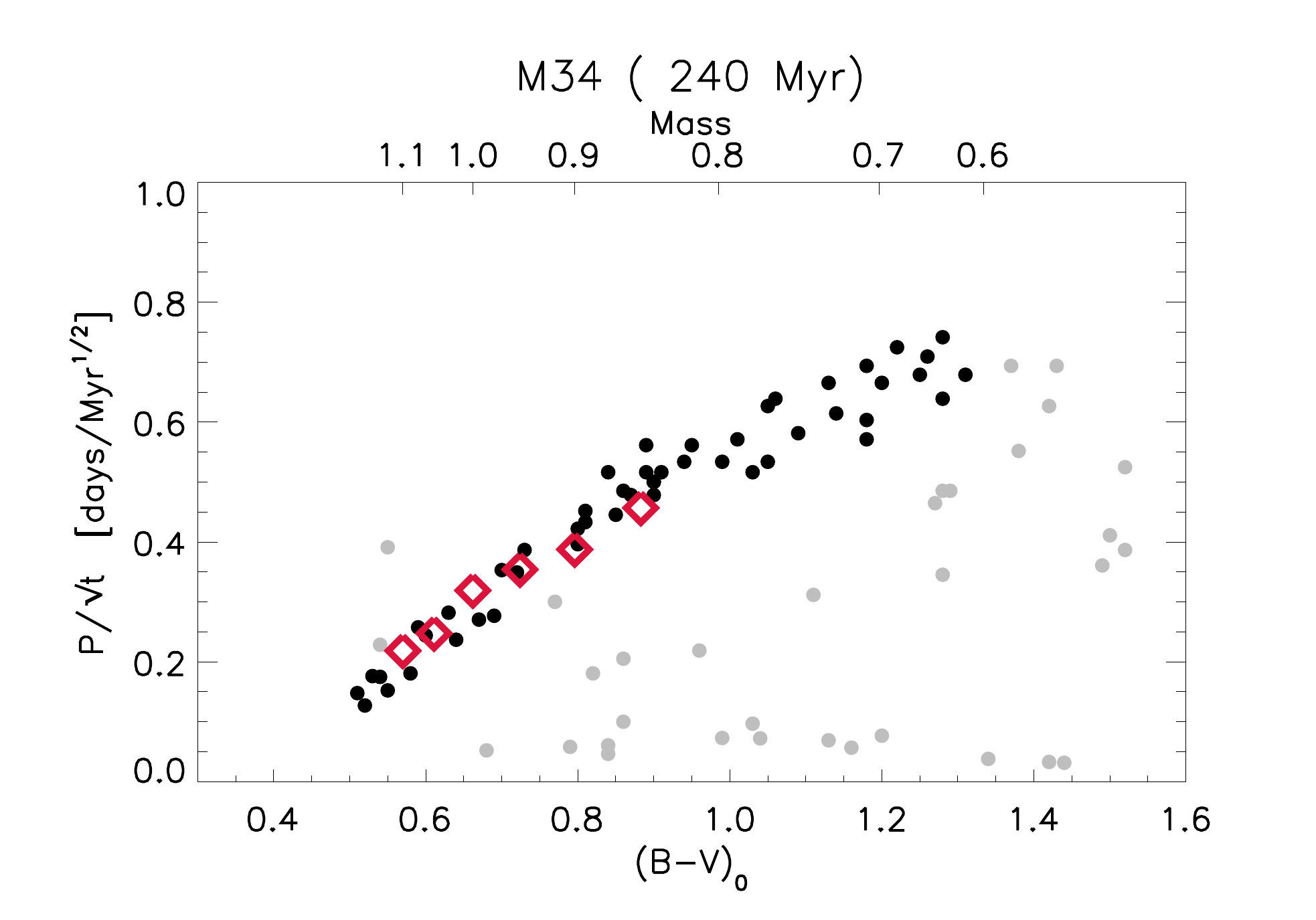}
\includegraphics[width=0.44\textwidth]{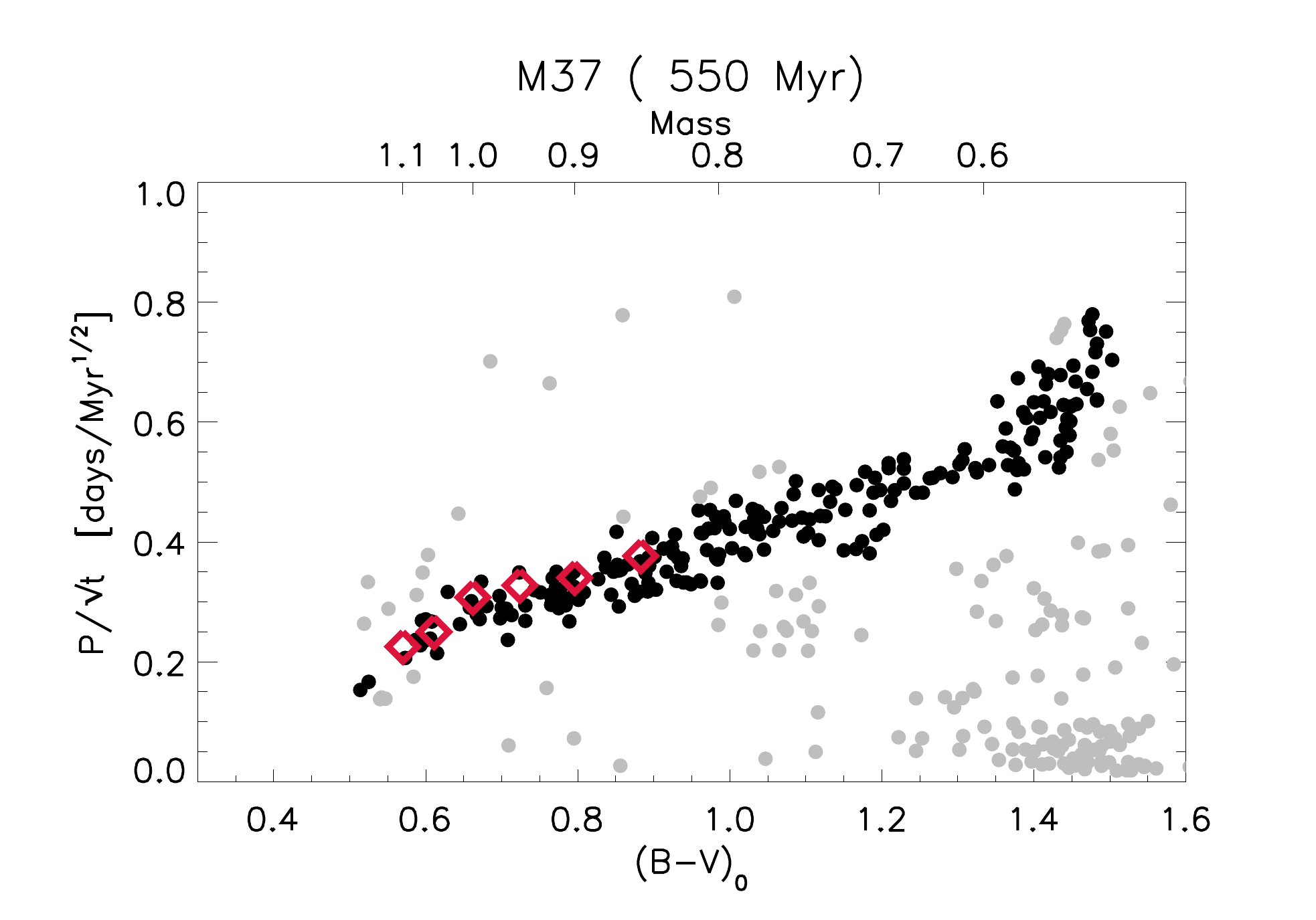}
\includegraphics[width=0.44\textwidth]{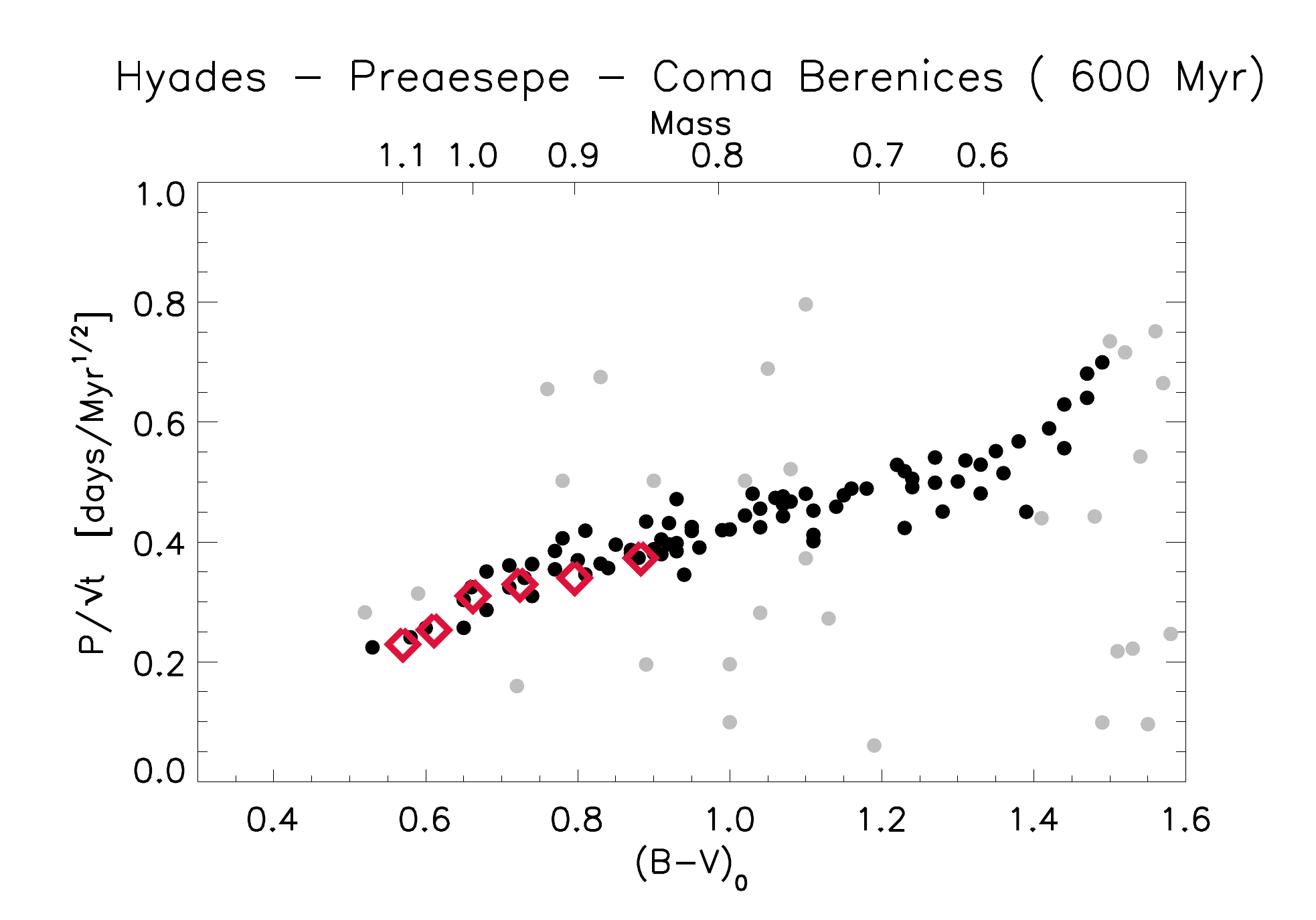}
\includegraphics[width=0.44\textwidth]{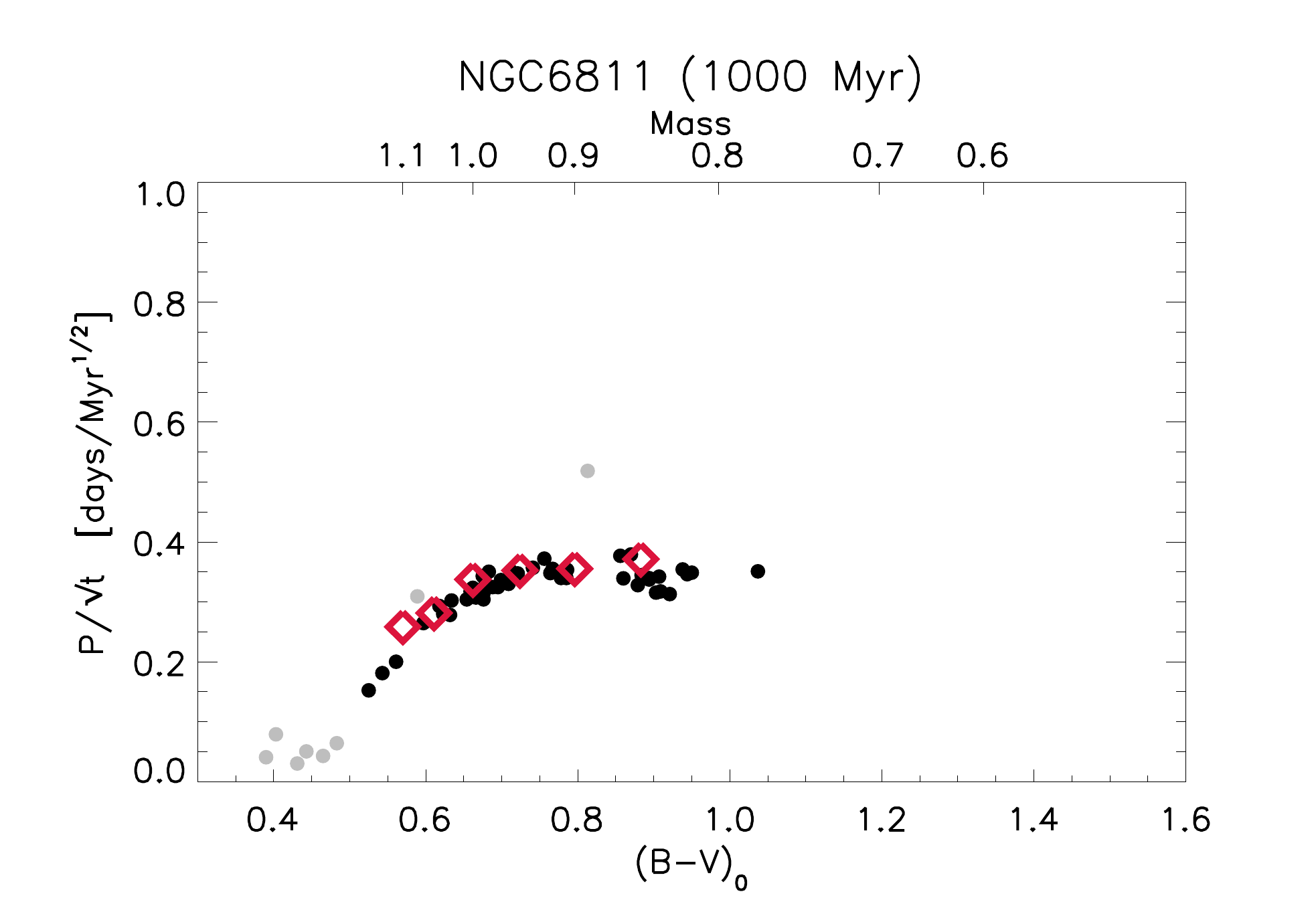}
\includegraphics[width=0.44\textwidth]{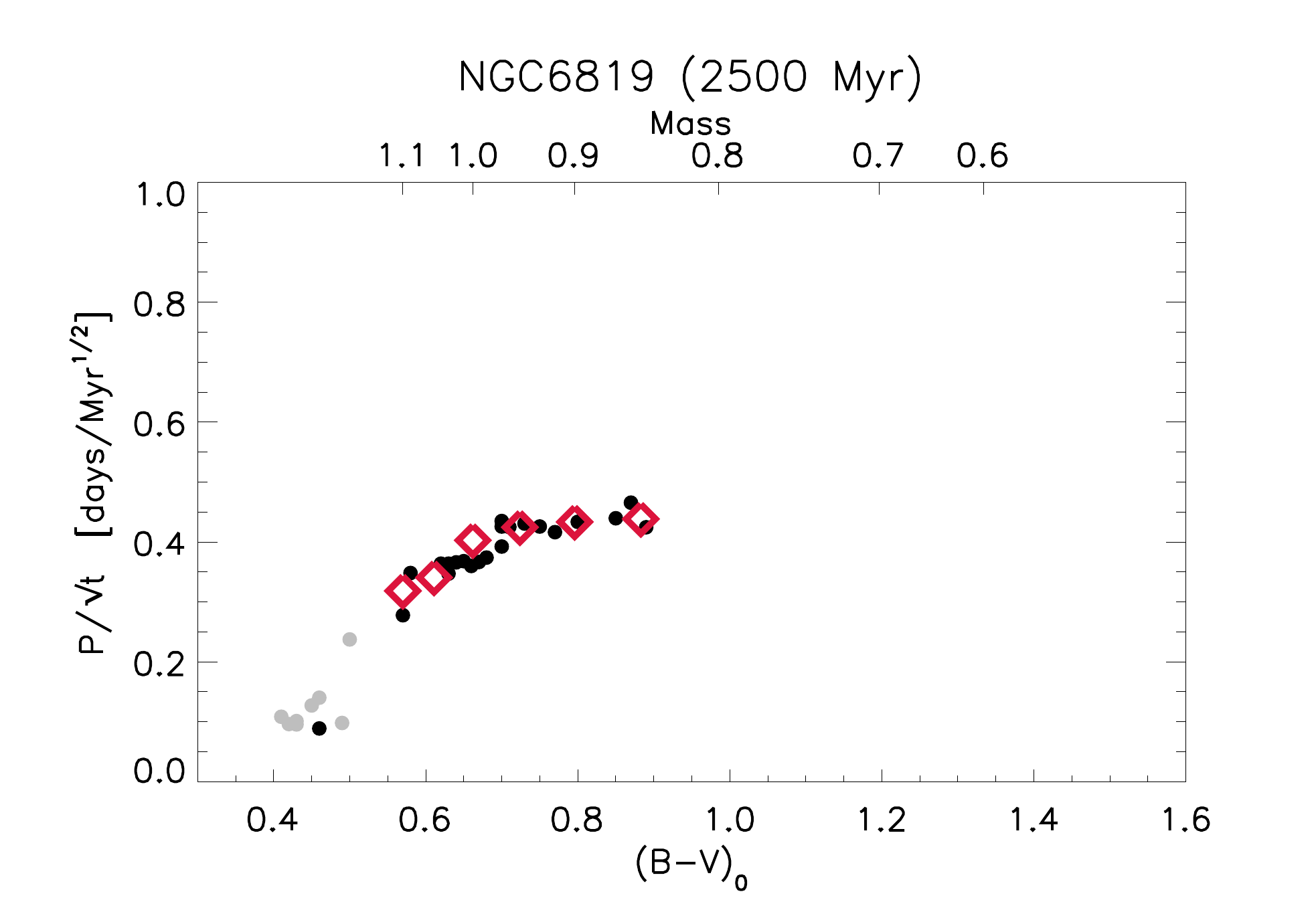}
\caption{Two-zone model MCMC fitting of the slow-rotator sequence (red diamonds) on the period-color diagram (black dots for the slow-rotator sequence, grey for others).
Parameters from model KB are used, adopting $\langle K_w \rangle = 4.5$  and correcting $\tau_{\rm cp}$ for likely biases according to the power-law fitting shown in Fig.\,\ref{fig:tauc_vs_mass}.
}
\label{fig:GyA_periods_iso}
\end{center}
\end{figure*}

The results of the fitting of the MCMC parameters are reported in Table\,\ref{tab:mcmc_results}, Fig.\,\ref{fig:Kw_vs_mass} and \ref{fig:tauc_vs_mass}.
Fig.\,\ref{fig:Kw_vs_mass} shows that the $K_w$ vs. $M$ relation obtained for the KB models produces the lowest linear Pearson correlation coefficient, compatible, within the uncertainties, with no correlation.
It is, indeed, remarkable that at $t\approx 0.55$\,Gyr the slow-rotator sequence follows the shape predicted by \cite{Barnes:2010} for the asymptotic slow-rotator sequence (Eq.\,\ref{eq:I-limit}) with an appropriate rescaling of the free parameter $k_I$.
In a solid-body rotation regime, this shape comparison would manifestly imply the mass scaling proportional to $I_* \tau$, implemented in model KB.
Although stars at 0.55\,Gyr have not yet reached this regime, our fit supports the validity of such mass scaling, at least in the mass range 0.85--1.1\,$M_{\odot}$.  
The confirmation of the general validity of such an empirical mass scaling awaits the acquisition of period data for $M<0.8\, M_{\odot}$ at ages well beyond the current 0.6\,Gyr limit.

Model MB also produces a significantly lower $K_w$ vs. $M$ correlation than the KA, KS, KK, and GB model.
However, the $K_w$ rise at $M > 1.0\, M_{\odot}$ indicates that the mass dependence of the MB wind braking law is less accurate than model KB at larger mass.

Model KA, i.e. the original \cite{Kawaler:1988} model with no saturation, tends to underestimate the wind braking at lower mass with respect to higher mass (or vice versa). 
Our fit compensates for this behavior with a systematic variation of $K_w$ with mass. 
A saturation threshold constant in mass, as in model KS, helps to correct this behavior, but only by a negligible amount for $M>1\,M_{\odot}$.
A mass-dependent saturation threshold, as in Eq.\,\eqref{eq:Krishna}, is somewhat more effective, but a clear $K_w$ vs. $M$ correlation remains.

The fitted values of $K_1$ for model GB are very close to the one given in \cite{Gallet_Bouvier:2013}, which derives from the original simulation of \cite{Matt_etal:2012a}, but show a clear correlation with $M$.
A strong correlation of the fitted $K_1$ with mass (using percentiles of the whole period sample) was also found by \cite{Gallet_Bouvier:2015}, who interpreted it as due to a change in magnetic configuration with mass.

A comparison of the wind braking mass dependence of models KA, KB, and MB in the range $0.6 \le M/M_{\odot} \le 1.1$ is shown in Fig.\,\ref{fig:mass_dependence}.
Comparing the latter with Fig.\,\ref{fig:Kw_vs_mass}, we deduce that the slope of the mass dependence of both models KB and MB is compatible with the observations in the range $0.85 \le M/M_{\odot} \le 1.0$.
For $M > 1.0\, M_{\odot}$, however, the slope of model MB is too steep and the observations favor a slope more similar to that of model KB.
Note, furthermore, that the effective mass dependence of model MB depends on the saturation regime, although this has only a minor impact on the $K_w$ fit, as in model KK.

\begin{table}
\caption{Periods derived from model KB with $\langle K_w \rangle = 4.5$ and $\tau_{\rm cp}$ corrected using the power-law fitting to the MCMC results (Fig.\,\ref{fig:tauc_vs_mass}) for a logarithmic-spaced age grid.}
\begin{center}
\begin{tabular}{rrrrrrr}
\hline
    Age  &   \multicolumn{6}{c}{$M/M_{\odot}$} \\
    (Myr)&   1.10  &  1.05  &  1.00  &  0.95  &  0.90  &  0.85 \\
\hline
    100.  &  2.65 &   3.04  &  4.04  &  4.62  &  5.22  &  6.38 \\
    125.  &  2.78 &   3.18  &  4.18  &  4.76  &  5.36  &  6.51 \\
    160.  &  2.95 &   3.37  &  4.41  &  4.98  &  5.55  &  6.68 \\ 
    200.  &  3.17 &   3.59  &  4.67  &  5.23  &  5.77  &  6.87 \\
    250.  &  3.44 &   3.90  &  5.01  &  5.55  &  6.06  &  7.13 \\
    320.  &  3.85 &   4.33  &  5.49  &  6.02  &  6.49  &  7.51 \\
    400.  &  4.33 &   4.85  &  6.08  &  6.59  &  6.99  &  7.95 \\
    500.  &  4.97 &   5.52  &  6.83  &  7.32  &  7.64  &  8.54 \\
    630.  &  5.80 &   6.41  &  7.82  &  8.30  &  8.54  &  9.33 \\
    800.  &  6.91 &   7.57  &  9.15  &  9.60  &  9.77  & 10.42 \\
   1000.  &  8.17 &   8.90  & 10.67  & 11.15  & 11.24  & 11.75 \\
   1250.  &  9.67 &  10.48  & 12.49  & 13.03  & 13.12  & 13.47 \\
   1600.  & 11.62 &  12.52  & 14.86  & 15.54  & 15.66  & 15.90 \\
   2000.  & 13.64 &  14.63  & 17.34  & 18.19  & 18.45  & 18.64 \\
   2500.  & 15.93 &  17.03  & 20.14  & 21.23  & 21.68  & 21.92 \\
   3200.  & 18.85 &  20.06  & 23.68  & 25.05  & 25.78  & 26.20 \\
   4000.  & 21.98 &  23.21  & 27.32  & 28.96  & 29.98  & 30.66 \\
   5000.  & 26.17 &  26.96  & 31.50  & 33.40  & 34.72  & 35.70 \\
\hline
\end{tabular}
\end{center}
\label{tab:periods_isochrones}
\end{table}

The resulting $\tau_{\rm cp}$ is largely independent of the model adopted (Fig.\,\ref{fig:tauc_vs_mass}).
The largest difference is between model GB, which is the only one with a wind braking law not strictly proportional to $\Omega^3$, and all the others.
Note that our $\tau_{\rm cp}$ derived from model GB is very close to that derived by \cite{Gallet_Bouvier:2015} for the 90th percentile of the whole period distribution.
As discussed below, a different $\Omega$ dependence results in a different angular momentum storage in the radiative interior, and therefore a different coupling timescale, but $\tau_{\rm cp}$ remains a steep function of mass in all cases.
Adopting model KB with an average $\langle K_w \rangle = 4.5$, we obtain a $\tau_{\rm cp}$ vs. $M$ relationship
which is well-represented by the power-law scaling:
\begin{equation}
\tau_{\rm cp} \propto M^{-7.28}.
\end{equation}
In Fig.\,\ref{fig:tauc_vs_mass} this relationships is compared with the $\tau_{\rm cp}$ obtained from each of the models tested in the MCMC fit.

The evolution of the envelope and core angular velocities is plotted in Fig.\,\ref{fig:tracks}.
For clarity, only the results of models KB, MB, and GB are shown. 
Since the mass dependence has been corrected by the $K_w$ fit, all models with a wind braking law proportional to $\Omega^3$ produce essentially the same $\Omega_e$ and $\Omega_c$ evolution.

In the $0.1$--$2.5$ Gyr age range, the $\Omega_e$ evolution deduced from model GB also overlaps with that deduced from all the other models.
Differences with the other models become significant after $2.5$ Gyr, but remain small even at the solar age (see Fig.\,\ref{fig:tracks}).
Model GB model predicts, however, a larger angular momentum storage in the core at early ages with respect to the other models.

The comparison of Figs.\,\ref{fig:GyA_periods} and \ref{fig:tracks} explains the behavior of the slow-rotator sequence in the $P/\sqrt{t}$ vs. $(B-V)$ diagram.
For $M < 1.0\,M_{\odot}$ at 0.1\,Gyr, $\Omega_e$ is well below the 
\begin{equation}
\label{eq:omega_scaling}
\Omega_e(t) = \Omega_{\rm obs}(t_0) \sqrt{\frac{t_0}{t}} 
\end{equation}
relationship with $t_0=0.55$\,Gyr, and $\Omega_{\rm obs}$ the corresponding value from the non parametric fit (Sect.\,\ref{sec:non-parametric}).
The evolution proceeds by approaching such relationship until $t=0.55$ Gyr, then $\Omega_e$ decreases more steeply than Eq.\,\eqref{eq:omega_scaling}.
In the time-frames of Fig.\,\ref{fig:GyA_periods}, this corresponds to a slow-rotator sequence approaching
the M37 sequence between $0.1$ and $0.55$\,Gyr, then departing from it between $0.55$ and $2.5$\,Gyr towards increasing $P/\sqrt{t}$.
The behaviour for $M \ge 1 M_{\odot}$ is qualitatively similar, but, in this case, $\Omega_e$ remains quite close to Eq.\,\eqref{eq:omega_scaling} until $0.55$\,Gyr, then decreases significantly faster than Eq.\,\eqref{eq:omega_scaling} afterwards.
This explains why the slow-rotator sequence for $M \ge 1 M_{\odot}$ remains close to the M37 sequence in the $P/\sqrt{t}$ vs. $(B-V)$ diagram until $0.55$\,Gyr, and then departs from it again from 0.55 to 2.5\,Gyr toward higher $P/\sqrt{t}$ values.

The comparison between $\Omega_e$ and $\Omega_c$ shows that a quasi-solid body rotation regime, and therefore a regime in which the wind braking dominates the rotational evolution, is achieved well after 1\,Gyr.
Furthermore, the comparison between $\Omega_e$ and Eq.\,\eqref{eq:omega_scaling} with $t_0=2.5$\,Gyr shows that a quasi-Skumanich evolution is achieved well after 1 Gyr.
This implies that in order to verify a departure from the $\Omega^3$ proportionality of the wind braking law, more data at ages well above 1\,Gyr are needed. 

\begin{figure}[ht]
\begin{center}
\includegraphics[width=0.48\textwidth]{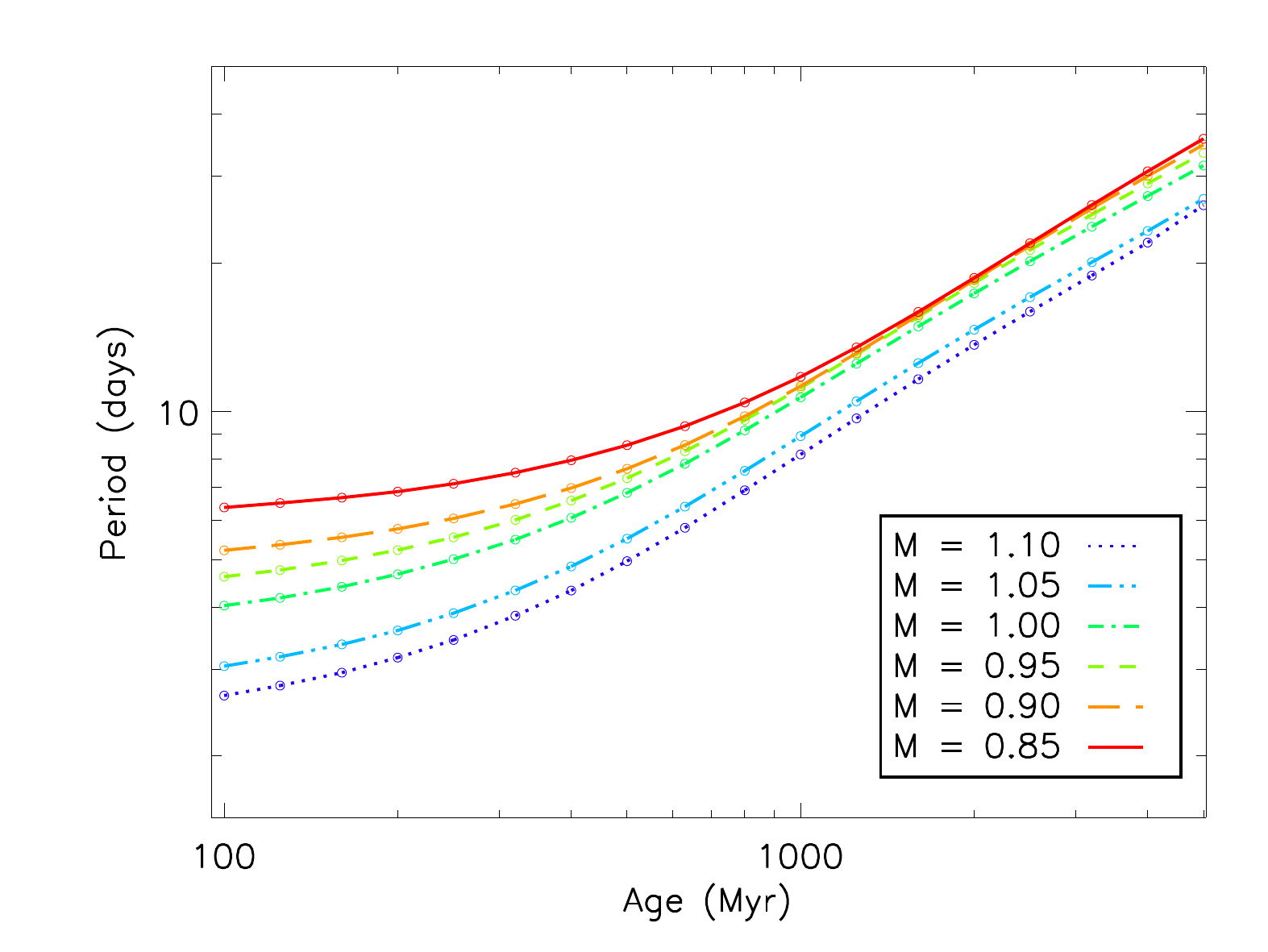}
\caption{Period evolutionary tracks calculated with model KB adopting $\langle K_w \rangle = 4.5$ and correcting $\tau_{\rm cp}$ for likely biases according to the power-law fitting shown in (Fig.\,\ref{fig:tauc_vs_mass}).
}
\label{fig:piso_vs_age}
\end{center}
\end{figure}

Period isochrones in the period-color diagrams are calculated with model KB adopting $\langle K_w \rangle = 4.5$ and correcting $\tau_{\rm cp}$ for likely biases according to the power-law fitting shown in Fig.\,\ref{fig:tauc_vs_mass}.
These isochrones are essentially equivalent to those computed using any of the wind braking models tested after correcting the mass dependence through the fitted $K_w$ and correcting $\tau_{\rm cp}$ in the same way.  
The comparison with observations, shown in Fig.\,\ref{fig:GyA_periods_iso}, outlines the observational origin of the small discrepancies found at the ages of the Pleiades and M35 for $M \ge 1.0\, M_{\odot}$, where photometry and reddening uncertainties, particularly crucial for a cluster like M35, affect more significantly the results.
Fig.\,\ref{fig:GyA_periods_iso} also shows that our model reproduces satisfactorily the period distribution in NGC6811 as well, even if it had been originally excluded from the MCMC fit.
Even the reconstructed period at $M=0.85\,M_{\odot}$, where the distribution apparently bends towards lower periods, lies on the upper envelope of the observed distribution and it is therefore compatible, in a statistical sense, with the observed distribution.
Overall, however, the evolution of the slow-rotator sequence is reproduced with unprecedented accuracy.

Period isochrones recomputed for a logarithmically spaced age grid are reported in Table\,\ref{tab:periods_isochrones} and Fig.\,\ref{fig:piso_vs_age}.
Interpolation of Table\,\ref{tab:periods_isochrones} provides a tool for inferring stellar ages of solar-like main-sequence stars from their mass and rotational period, effectively representing an alternative gyro-chronology relationship that takes the physics of the two-zone model for the stellar angular momentum evolution into account.

\section{Conclusions}
\label{sec:conclusions}

Using a two-zone model MCMC fitting to the rotational periods vs color data from selected open clusters, we have compared new proposals for the wind braking law with the classical \cite{Kawaler:1988} one modified by \cite{Chaboyer_etal:1995}.
In our tests, we coupled the \cite{Kawaler:1988} braking law with different hypotheses regarding the saturation regime, including that of \cite{Krishnamurthi_etal:1997}.
The wind braking laws tested also included that of \cite{Gallet_Bouvier:2013}, one derived from the work of \cite{Barnes_Kim:2010} and \cite{Barnes:2010}, and the new proposal by \cite{Matt_etal:2015}.

The MCMC fitting has been performed on the slow-rotator sequence between $0.1$ and $2.5$ Gyr.
We proposed a new quantitative criterion for identifying the slow-rotator sequence based on the symmetry of the data distribution around the density peak in the period-color diagram.
Following this criterion, it is possible to assign a width of the period distribution around the density peak using the standard deviation.
The value of the peak density as a function of colors (mass) is estimated using a non-parametric fit and, together with the standard deviation, constitute the data used in the MCMC fitting and in the definition of the likelihood function.

The main idea behind this work was to see how accurately a two-zone model with parameters constant in time can describe the rotational evolution on the slow-rotator sequence.
In fact, it is evident that the slow-rotator sequence evolves in a regular and predictable way and it is then natural to explore the extent to which current models can reproduce this sort evolution.

We find that the difficulties in reconciling the observed behavior of the slow-rotator sequence evolution between 0.1 and 0.6 Gyr with the original \cite{Barnes:2003} idea of a factorization of time and mass dependence \citep{Meibom_etal:2009,Meibom_etal:2011} can be simply attributed to a transfer of angular momentum from the radiative core to the convective envelope in the early slow-rotator sequence evolution.
The two-zone model MCMC fitting leads to a robust (largely independent of the wind braking model adopted) estimate of the core-envelope coupling timescale $\tau_{\rm cp}$ as a function of mass in the range $0.85 < M/M_{\odot} < 1.10$.
In this mass range we find that $\tau_{\rm cp}$ scales as $M^{-7.28}$ on the slow-rotator sequence.

The importance of the core angular momentum storage and transfer to the convective envelope, in which enhanced viscosity that is due to (magneto-) hydrodynamical instabilities is expected to play an important role, has been outlined by many authors (see, e.g. references in Sect.\,\ref{sec:tzm}).
In this work we derive, for the first time, the empirical mass dependence of the core-envelope  coupling timescale in the slow-rotator regime. 
Our empirical mass dependence of $\tau_{\rm cp}$ represents a useful term of comparison for the theoretical description of angular momentum transport in stellar interiors.
By focusing on the slow-rotator sequence, in which the TZM with constant parameters, as described in Sect.\,\ref{sec:tzm}, is expected to capture the essential physics of the angular momentum evolution, we avoid the complications that arise in dealing with the fast-rotator regime, in which the TZM with constant parameters may break down \citep[e.g.][]{Brown:2014}.
The validity of the TZM for the slow-rotator regime is confirmed by the successful fit to the cluster data currently available.
The wind braking law, for which the empirical dependence on stellar mass and angular velocity still gives better results than purely theoretical ones \citep[see e.g.][]{Matt_etal:2015}, remains one of the critical aspects of this modeling.

Our results outline and explain the invalidity of a factorization like the one expressed by Eq.\,\eqref{eq:factorisation}, on which most of the proposed gyro-chronology relationships are based, at least for the early (mass-dependent) evolution of the slow-rotator sequence.
On the other hand, our results mean that the search for gyro-chronology relationships like the one recently proposed by \cite{Kovacs:2015}, which is not based on known physics, are not justified.
\cite{Kovacs:2015} proposal was motivated in fact by the disagreement between gyro-chronology relationships based on Eq.\,\eqref{eq:factorisation} and stellar evolution isochrone fitting, but the ages assumed for the clusters included in our analysis, which do not differ significantly from the isochrone ages assumed in \cite{Kovacs:2015}, are fully compatible with our modeling of the rotational evolution of the slow-rotator sequence.

With a suitable choice of the free parameters, we tested the mass dependence of the models under scrutiny.
We found that the mass scaling of the wind braking law implied by the models of \cite{Barnes:2010} and \cite{Barnes_Kim:2010} is the most consistent with the data, supporting the proposal that the convective turnover timescale $\tau$ is a key parameter in such a mass dependence.
A definitive confirmation of the general validity of such mass dependence, however, would require period data of older cluster and lower stellar mass than is available to date.
The mass scaling of the recent model by \cite{Matt_etal:2015} is found consistent with the data for $0.85 \le M \le 1.00\,M_{\odot}$, but it decreases too steeply with mass for $M > 1.00\,M_{\odot}$.
All the other models considered show a clear residual correlation with mass, indicating an incorrect mass scaling.

Small deviations from the $\Omega^3$ dependence of the wind braking law, such as those predicted by the \cite{Gallet_Bouvier:2013} model, are also compatible with observations in the 0.1--2.5\,Gyr range.
In fact, quasi-solid body rotation, the regime in which the wind braking dominates the angular momentum evolution, is achieved after 1--2\,Gyr, depending on mass, so that small deviations from the $\Omega^3$ proportionality can be compensated for by a different angular momentum storage in the stellar core to produce, essentially, the same $\Omega_e$ evolution.
After correcting the mass scaling, differences between the \cite{Gallet_Bouvier:2013} model and models based on the $\Omega^3$ proportionality becomes not negligible after 2.5\,Gyr, although they remain small even at the age of the Sun.
A corollary of this is that verifying small deviations from the $\Omega^3$ proportionality would require data for clusters much older than 2.5\,Gyr and that, in the 2--5\,Gyr age range, deviations from the Skumanich law for $M \approx 1\,M_{\odot}$ are expected to be small.

From our MCMC model fitting to the data, we derive period isochrones and tracks from 0.1 to 5.0\,Gyr valid for stars with $0.85 \le M \le 1.10\,M_{\odot}$.
These are essentially independent of any assumption regarding the mass scaling of the wind braking law and compatible with small deviations from the $\Omega^3$ proportionality.
By interpolating between the grid point, these tracks can be used to infer stellar ages from their mass and rotational period, providing alternative gyro-chronology relationships that take the physics of the two-zone model into account.


\begin{acknowledgements}
We are grateful to an anonymous referee, whose comments have contributed to improving the paper.
ACL thanks the Cosmic Magnetic Fields Branch of the Leibniz-Institut f\"ur Astrophysik Potsdam (AIP) for their kind hospitality.
ACL acknowledges the support received from the European Science Foundation (ESF) for the activity entitled 'Gaia Research for European Astronomy Training'.
FS acknowledges support from the Leibniz Institute for Astrophysics Potsdam (AIP) through the Karl Schwarzschild Postdoctoral Fellowship.
\end{acknowledgements}

\bibliographystyle{aa}
\bibliography{SlowRotatorsYoung}

\end{document}